\documentstyle[epsf]{mn} 

\onecolumn
 
\def\kms{km ${\rm s}^{-1}$}
\def\gx339{GX~339$-$4} 
\def\gro1655{GRO~J1655$-$40}
\def\GS{\lower.5ex\hbox{$\buildrel>\over\sim$}}
\def\LS{\lower.5ex\hbox{$\buildrel<\over\sim$}} 

\title[Optical Spectroscopy of GX~339$-$4 II]
{Optical Spectroscopy of GX~339$-$4 during the High-Soft and 
   Low-Hard States II: Line Ionisation and Emission Region } 

\author[K. Wu et al.]{ Kinwah Wu$^1$, Roberto Soria$^{2}$,
   Richard W. Hunstead$^3$ and Helen M. Johnston$^4$
 \\
$^1$ Research Centre for Theoretical Astrophysics, School of Physics,
   University of Sydney, NSW 2006, Australia; \\ 
\ \  kinwah@physics.usyd.edu.au \\ 
$^2$ Mullard Space Science Laboratory, University College London, 
   Holmbury St Mary, Dorking, \\ 
\ \  RH5 6NT, UK; rs1@mssl.ucl.ac.uk \\
$^3$ Astrophysics Department, School of Physics, University of Sydney, 
   NSW 2006, Australia; rwh@physics.usyd.edu.au \\ 
$^4$ Anglo-Australian Observatory, P.~O.~Box 196, Epping, NSW 1710, 
   Australia; hmj@aaoepp.aao.gov.au  }  
 
\date{Received: } 
 
\begin{document}

\maketitle 

\begin{abstract}  
We have carried out observations of the X-ray transient \gx339 during 
its high-soft and low-hard X-ray spectral states. Our high-resolution 
spectroscopic observation in 1999 April suggests that the H$\alpha$ line 
has a single-peaked profile in the low-hard state as speculated in our 
previous paper. The He\,{\scriptsize II}~$\lambda$\,4686 line, however, 
has a double-peaked profile in both the high-soft and low-hard states. 
This suggests that the line-emission mechanism is different in the 
two states. Our interpretation is that double-peaked lines are emitted 
from a temperature-inversion layer on the accretion-disk surface 
when it is irradiatively heated by soft X-rays. Single-peaked lines 
may be emitted from outflow/wind matter driven by hard X-ray heating. 
We have constructed a simple plane-parallel model and we use it to 
illustrate that a temperature-inversion layer can be formed at the disk 
surface under X-ray illumination. We also discuss the conditions 
required for the formation of temperature inversion and line emission. 
Based on the velocity separations measured for the double-peaked lines 
in the high-soft state, we propose that \gx339 is a low-inclination 
binary system. The orbital inclination is about $15^\circ$ if the 
orbital period is 14.8~hours. 

\end{abstract}


\section{Introduction}   

\gx339 is an X-ray transient (Markert et\,al.\ 1973; 
Harmon et\,al.\ 1994), and is classified as a black hole 
candidate (BHC) because of its short-term X-ray and optical 
variability (Makishima et\,al.\ 1986), its transitions between 
high-soft and low-hard X-ray spectral states (Markert et\,al.\ 1973), 
and the extended high-energy power-law tail in its X-ray spectrum 
(Rubin et\,al. 1998). It is one of the few BHCs that has shown all 
four X-ray spectral states: {\it off}, {\it low-hard}, {\it high-soft} 
and {\it ultra-high}. During the off state, \gx339 has very weak and 
hard X-ray emission. Its optical counterpart is faint, with 
$V \sim 19 - 21$. In the low-hard state, it emits very strong 
hard X-rays. The spectrum is an extended power-law with a photon 
index $\sim 1.5$. The 2$-$10~keV X-ray flux is, however, weak 
($\sim 0.4 \times 10^{-9}\;{\rm erg\,cm^{-2}\,s^{-1}}$). 
The optical brightness of the system is $V \sim 16 - 17$.  
In the high-soft state, the X-ray spectrum is characterised by a 
soft, thermal component. The 2$-$10~keV X-ray flux is about 20 times  
higher than in the low-hard state. The power-law tail is weak,  
with a photon index $\sim 2$. The optical brightness is similar to 
that in the low-hard state, with $V \sim 16-17$. 
In the ultra-high state, both the thermal and the power-law 
components in the X-ray spectrum are strong. The 2$-$10~keV 
X-ray flux is about 50 times the X-ray flux in the low state, and 
the photon index of the power law is $\sim 2.5$. On one occasion 
Mendez \& van der Klis (1997) reported that the source was in a 
state intermediate between the low-hard and the high-soft state. 
The 2$-$10~keV X-ray flux was about a factor of 5 below the X-ray 
flux in the high state.   

\gx339 is also a radio source (e.g.\ Hannikainen et\,al.\ 1998). It 
has a flat spectrum, and its image has shown jet-like features 
(Fender et\,al.\ 1997). The radio emission was found to be correlated 
with the hard (20---100~keV) X-ray flux, and was quenched during a 
high-soft state (Fender et al.\ 1999). 

Despite the fact that \gx339 is optically bright ($V\sim 16-17$) when 
it is X-ray active, it is not well studied in the optical bands. The 
mass function of the system has not yet been determined, and so the 
black hole candidacy is not verified in terms of the orbital dynamics. 
In a photometric observation during the off state, 
Callanan et\,al.\ (1992) detected a periodicity of 14.8~hr, and  
attributed it to the orbital period. A 14.8-hr orbital period would 
imply that the system has a late-type companion star with a mass 
$< 1.6$~M$_\odot$. The 14.8-hr period, however, was not seen in the 
photometric data that we collected in 1998 (Soria 2000; 
Soria, Wu \& Johnston 2000).    

In our previous paper (Soria, Wu \& Johnston 1999, hereafter Paper I) 
we reported part of the findings from our 1997 and 1998 spectroscopic 
observations. The system was in a low-hard state during the 1997 
observation and was in a high-soft state during the 1998 observations. 
We found that the emission lines generally show double-peaked profiles 
in the high-soft state. The H$\alpha$ line did not show resolved peaks 
in the medium-resolution (3 \AA) spectra that we obtained in the 
low-hard state. Because of the limited spectral coverage in our 1997 
observation, we were unable to determine the properties of 
the H$\beta$, He\,{\scriptsize II}~$\lambda$\,4686 and  
Bowen N\,{\scriptsize III} $\lambda \lambda$\,4641,4642 lines in the 
low-hard state. We therefore restricted our discussion to the 
profiles of the H$\alpha$ line and did not compare the properties of 
the high-ionisation lines in the two X-ray spectral states.

\gx339 was in its low-hard state again in early 1999, but it entered  
an X-ray quiescent phase in mid-April, with its hard X-ray flux below 
the detection limit of BATSE (from BATSE bright source reports) and 
its RXTE/ASM count rate consistent with zero (from quick-look results 
of the RXTE/ASM team). 
Its radio flux density in late June, measured by the ATCA, was more 
than 10 times weaker than that measured in an observation on 
April 14 (S.~Corbel, private communication). Our optical spectroscopic 
observation was carried out on April 12, about 5 days before it became 
undetectable by BATSE. High-resolution ($\sim 1$~\AA) spectra at the 
H$\alpha$ and the H$\beta$/He\,{\scriptsize II} regions were obtained. 
In this paper we report the results of the 1999 observation and compare 
them with those of our 1997 and 1998 observations. We discuss the 
profiles of the high- and low-ionisation lines in different X-ray 
spectral states and the implications for the line-formation mechanisms 
and the line-emission region in BHCs.  

\section{Optical Spectroscopy}  

Our spectroscopic observations were carried out with the 3.9-m 
Anglo-Australian Telescope (AAT) in 1997 May and 1999 April and with 
the ANU 2.3-m Telescope at Siding Spring Observatory in 1998 April and 
August. The observations covered two X-ray spectral states of the 
source: the low-hard state in 1997 and 1999, and the high-soft 
state in 1998. 

The observations are listed in Table 1.    
The technical details of the 1997 and 1998 observations were presented 
in Paper I. For the 1999 April observations, we used a 
Tektronix 1k$\times$1k CCD on the RGO spectrograph and a 1200V grating; 
the seeing was $<1$~arcsec. The spectra were centred at 6563~\AA\ 
(H$\alpha$ region) and at 4770~\AA\ 
(H$\beta$/He\,{\scriptsize II} region), and the spectral 
resolution was 1.3~\AA\ FWHM for both the H$\alpha$ and 
the H$\beta$/He\,{\scriptsize II} spectra. The total exposure time for 
the H$\alpha$ spectrum was 1200~s, and that for the 
H$\beta$/He\,{\scriptsize II} spectrum 600~s. Standard data reduction 
procedures were followed, using IRAF tasks, and wavelengths were 
calibrated using the Cu-Ar lamp spectra. 

\begin{table*}
 \centering
 \begin{minipage}{140mm}
  \caption{Log of our optical spectroscopic observations.}
  \begin{tabular}{lcccc}
\hline
\hline
UT date & epoch of observations & Wavelength range & Resolution  
& Telescope \\
& (HJD - 2450000)  &  (\AA) 
& (\AA\ FWHM) & \\
\hline
\hline
\multicolumn{4}{c}{low-hard X-ray spectral state } \\
\hline
1997 May 6 & 574.966--575.226 & 5355--6950 & 3 & AAT\\
1997 May 8 & 576.974--577.042 & 5355--6950 & 3 & AAT\\ 
1999 April 12 & 1281.287--1281.313 & 4455--5072 & 1.3 & AAT \\ 
              &       & 6282--6828 & 1.3 &   \\ 
\hline
\multicolumn{4}{c}{high-soft X-ray spectral state } \\
\hline
1998 April 28 & 931.953--932.318 & 4150--5115 & 1.3 & ANU 2.3~m\\
&&6200--7150 & 1.3 \\
1998 April 29 & 932.943--933.268 & 4150--5115 & 1.3 & ANU 2.3~m\\
&&6200--7150 & 1.3 \\
1998 April 30 & 934.106--934.233 & 4150--5115 & 1.3 & ANU 2.3~m\\
&&6200--7150 & 1.3 \\
1998 August 20 & 1045.864--1046.140 & 4150--5115 & 1.3 & ANU 2.3~m\\
&&6200--7150 & 1.3 \\
1998 August 23 & 1048.860--1049.103 & 4150--5115 & 2 & ANU 2.3~m\\
&&6200--7150 & 2 \\  
\hline
\end{tabular}
\end{minipage}
\end{table*} 

\begin{figure}
\vspace*{0.4cm} 
\begin{center}
\begin{tabular}{ccc}
  \epsfxsize=5.5cm \epsfbox{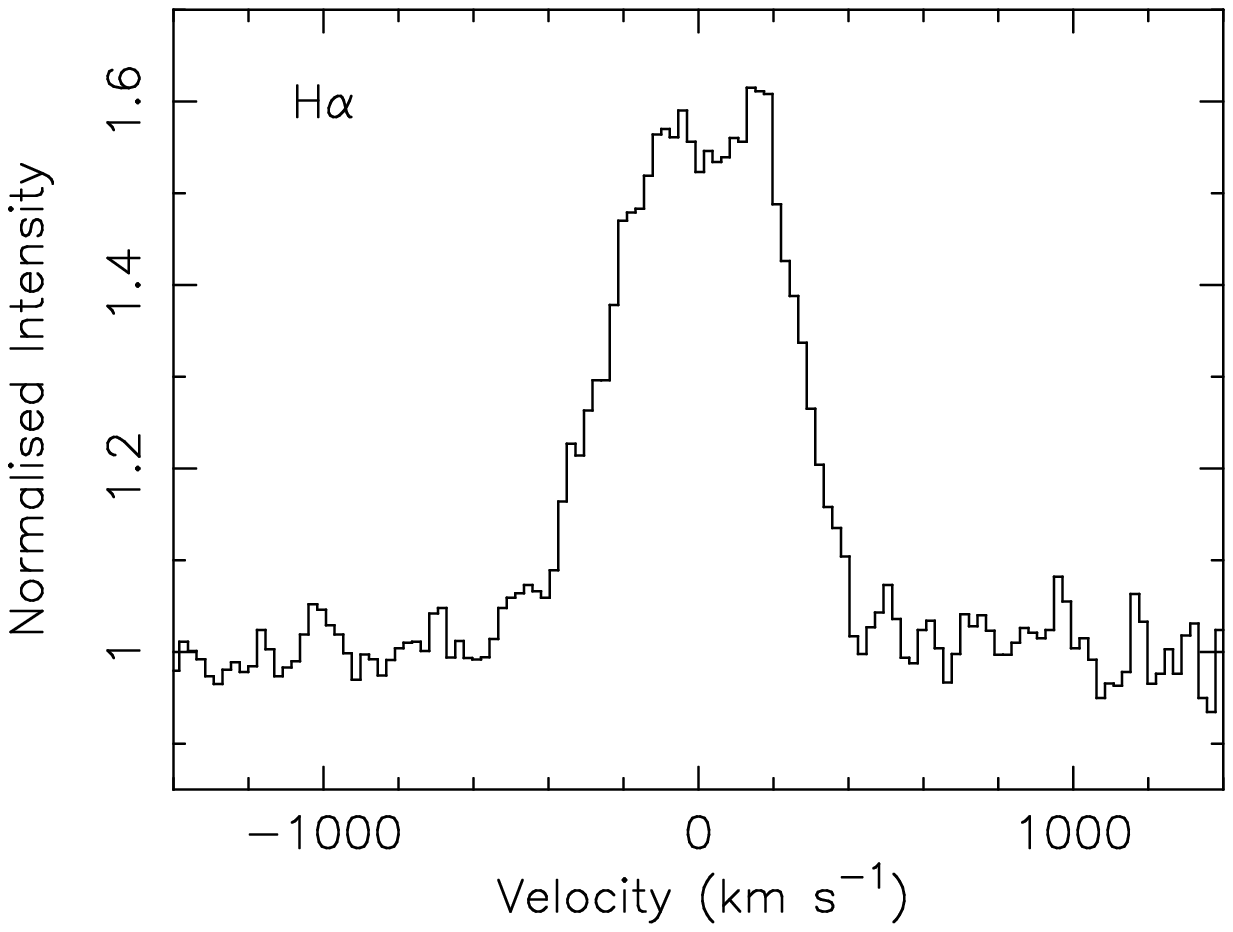} &
  \epsfxsize=5.5cm \epsfbox{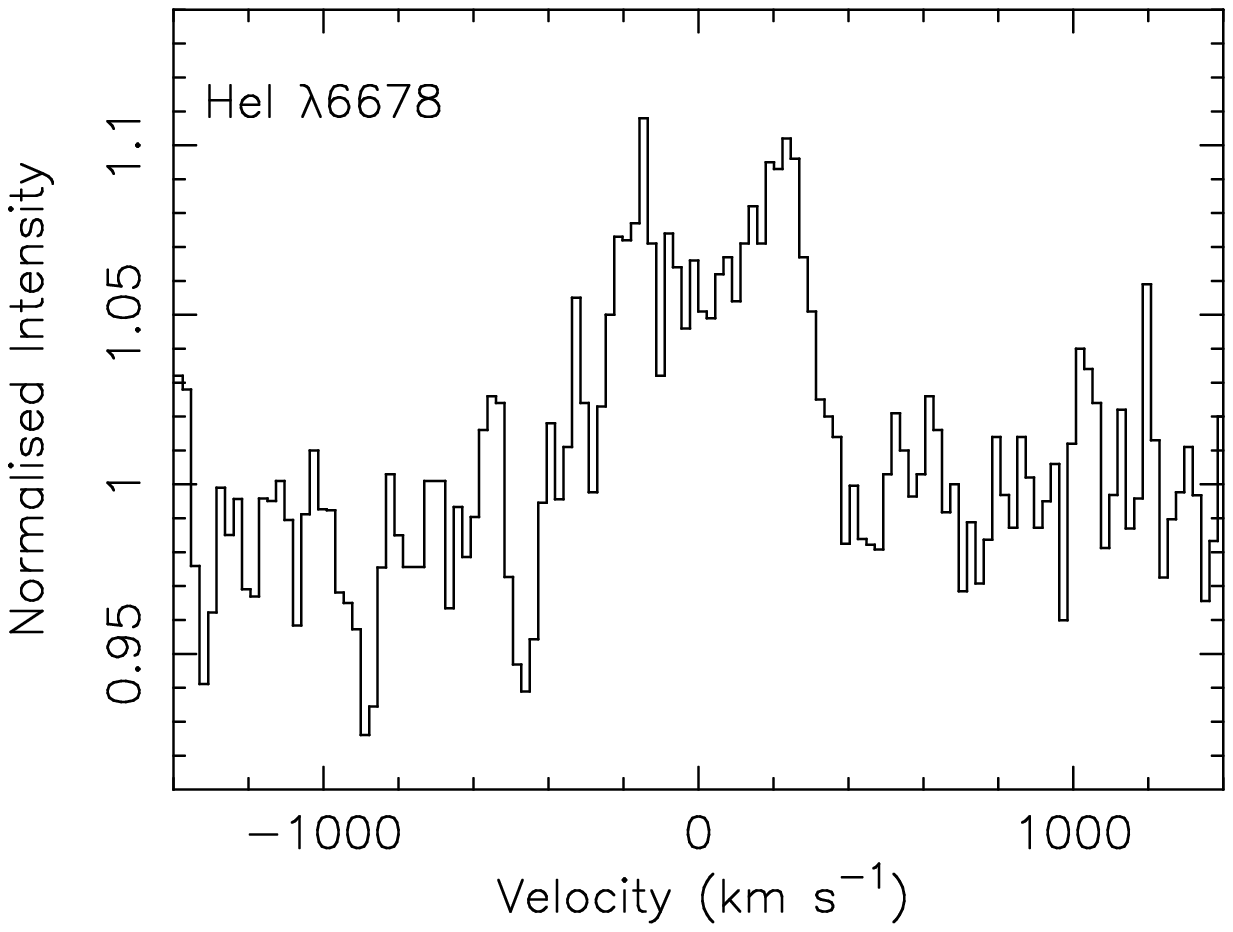} &
  \epsfxsize=5.5cm \epsfbox{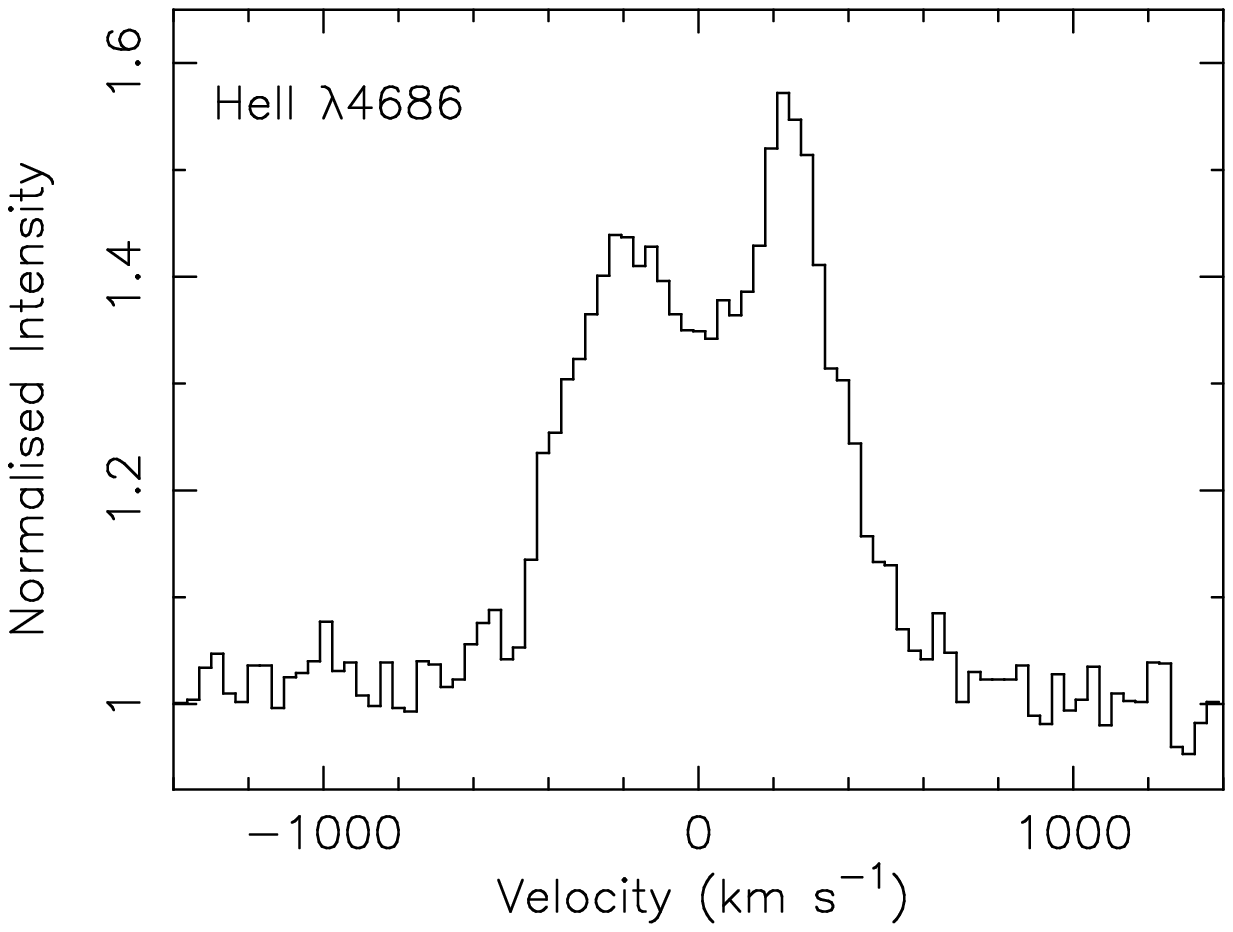} \\ 
  \epsfxsize=5.5cm \epsfbox{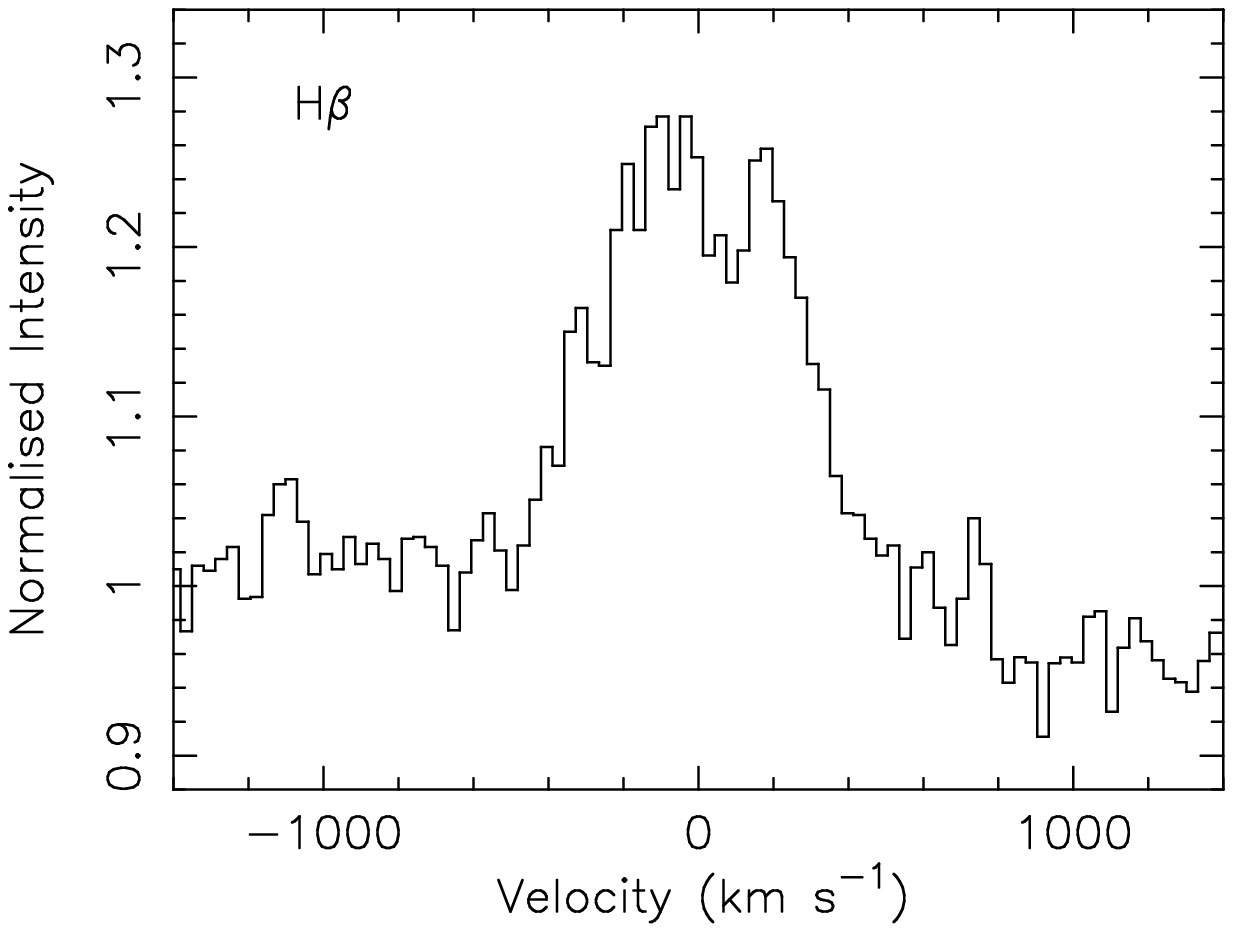} &
  \epsfxsize=5.5cm \epsfbox{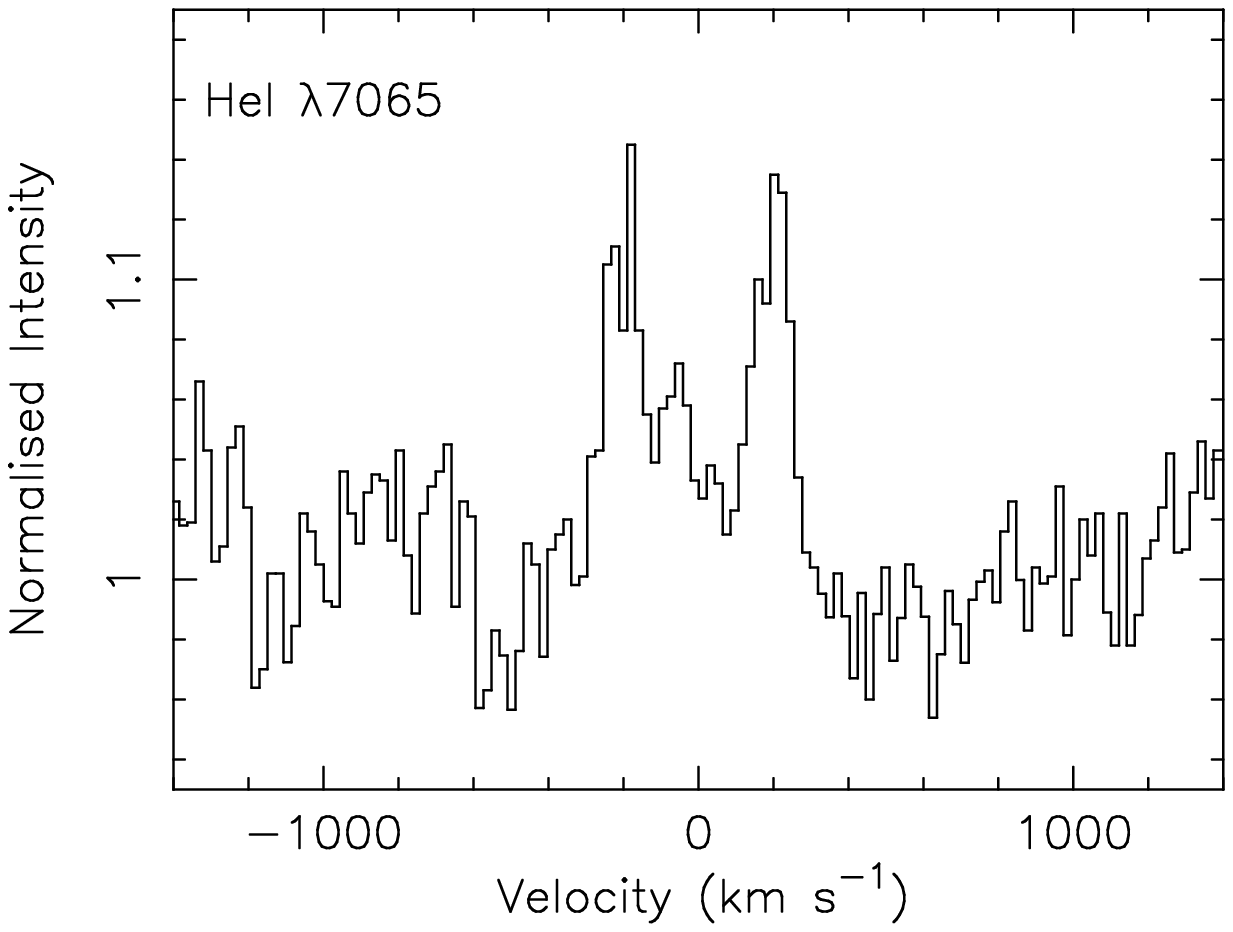} &
  \epsfxsize=5.5cm \epsfbox{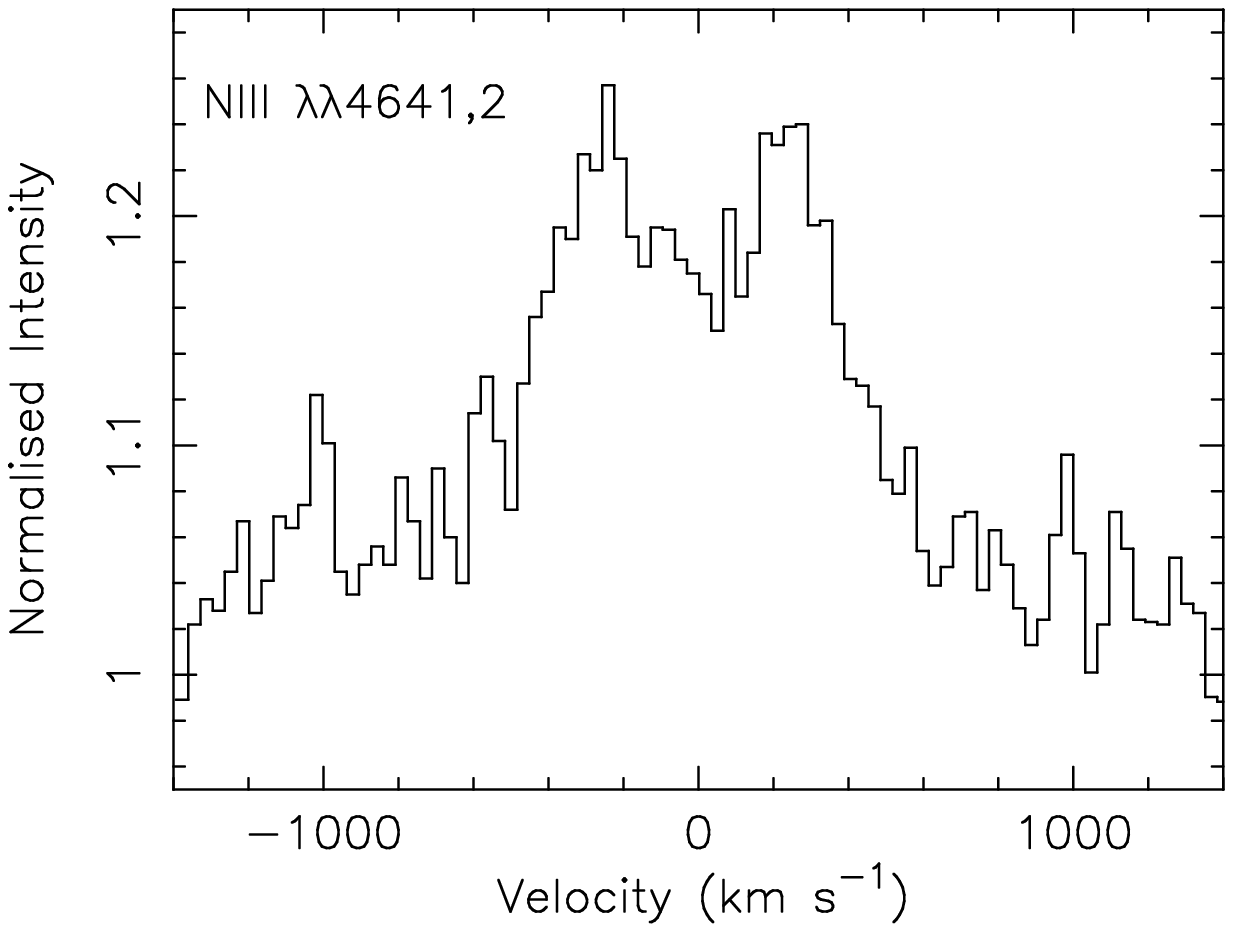}  
\end{tabular} 
\end{center}
\caption{The profiles of the H$\alpha$, H$\beta$,  
  He\,{\scriptsize I} $\lambda$\,6678, 
  He\,{\scriptsize I} $\lambda$\,7065, 
  He\,{\scriptsize II} $\lambda$\,4686 and 
  N\,{\scriptsize III} $\lambda \lambda$\,4641,4642 lines of \gx339 
  in 1998 April during a high-soft state. }
\end{figure} 

\section{Line peak-separation in the high-soft state}  

In Paper I we showed the H$\alpha$ and H$\beta$/He\,{\scriptsize II} 
spectra of \gx339 in 1998 April and August. The strong emission lines 
in the spectra are  H\,{\scriptsize I} Balmer, 
He\,{\scriptsize I} $\lambda$\,6678, 
He\,{\scriptsize I} $\lambda$\,7065, 
He\,{\scriptsize II} $\lambda$\,4686 
and Bowen N\,{\scriptsize III} $\lambda \lambda$\,4641,4642.  
Other detectable emission lines are 
He\,{\scriptsize I} $\lambda$\,4922,  
He\,{\scriptsize I} $\lambda$\,5876 and 
N\,{\scriptsize II}~$\lambda$\,6505. 
The spectra are very similar to those of other BHCs  
(e.g.\ \gro1655, see Soria, Wu \& Hunstead 2000) in their high-soft 
states. The spectra of these systems all show high-ionisation lines, 
and the emission lines are double-peaked.  

We show in Figure 1 the high-resolution profiles of the double-peaked 
H$\alpha$, H$\beta$, He\,{\scriptsize I} $\lambda$\,6678, 
He\,{\scriptsize I} $\lambda$\,7065, 
He\,{\scriptsize II} $\lambda$\,4686 and 
N\,{\scriptsize III} $\lambda \lambda$\,4641,4642 lines. 
The velocity separations of the peaks are $250\pm27$~\kms\ for 
H$\alpha$, $260\pm35$~\kms\ for H$\beta$, 
$380\pm38$~\kms\ for He\,{\scriptsize I} $\lambda$\,6678, 
$390\pm23$~\kms\ for He\,{\scriptsize I} $\lambda$\,7065,  
$480\pm27$~\kms\ for He\,{\scriptsize II} $\lambda$\,4686 and 
$490\pm66$~\kms\ for N\,{\scriptsize III} $\lambda \lambda$\,4641,4642. 
The velocity separations of H$\alpha$ and H$\beta$ are the same  
within the uncertainties of the measurements, and so are those of 
He\,{\scriptsize I} $\lambda$\,6678 and  
He\,{\scriptsize I} $\lambda$\,7065. The similarity of the velocity 
separations for He\,{\scriptsize II} $\lambda$\,4686 and  
N\,{\scriptsize III} $\lambda \lambda$\,4641,4642 implies that these 
lines are emitted at a similar distance from the compact object, 
consistent with the theory of Bowen fluorescence, which requires  
He\,{\scriptsize II} for the pumping mechanism 
(see McClintock, Canizares \& Tarter 1975; 
Kallman \& McCray 1980; Schachter, Filippenko \& Kahn 1989). 
Thus, we have observed a clear trend in which the velocity separations 
of the double-peaked lines increase with line ionisation, from the 
neutral hydrogen lines ($\approx 250$ \kms), to the neutral helium 
lines ($\approx 400$ \kms) and finally to the higher-ionisation 
(i.e.\ ionised helium and Bowen N\,{\scriptsize III}) lines 
($\approx 500$ \kms).  

The optical continuum of \gx339 was bright during our 1998 observations, 
suggesting the presence of an active, bright accretion disk. If the  
double-peaked lines are emitted from the accretion disk, as commonly 
interpreted, we can use the peak separations to infer the radial distance
of the emission regions from the compact object. Moreover, as 
higher-ionisation lines are emitted from hotter gases, we can also use 
the peak separations to map the surface temperature distribution of the  
accretion disk.  

Assuming that the lines are emitted from a 
geometrically thin, Keplerian accretion disk, we can 
use the line velocities to constrain the system parameters. 
In Figure 2 we plot the expected velocity separation as a function of 
distance from the compact object for systems with a compact object of 
mass 4.0, 7.0 and 10.0 M$_\odot$ and for various orbital inclinations.  
We also mark in the figure the expected orbital separation of the 
systems for various orbital periods, assuming that the companion star 
is a Roche-lobe-filling main-sequence star. As the size of the accretion 
disk cannot exceed the Roche lobe of the compact star, which is 
smaller than the orbital separation, we can easily see that the 
orbital inclination of \gx339 is well below 30$^\circ$ if the orbital 
period is 14.8~hr (Callanan et\,al.\ 1992). Moreover, the mass of the 
compact object is unlikely to exceed 10~M$_\odot$. However, the 
orbital inclination and the mass of the compact object may be higher if 
the orbital period is longer. If we adopt the BHC average mass of 
7~M$_\odot$ (see e.g.\ McClintock 1998) as the mass of the compact object 
in \gx339 and assume that the Balmer lines are emitted near the rim of 
the accretion disk and that the orbital period is 14.8~hr, then we deduce 
an orbital inclination of about 15$^\circ$ for the system. 

We note that no large velocity variations of the centres of the disk 
lines were detected in our 1998 April and August observations. This may 
be further evidence of a low orbital inclination. Additional evidence is 
provided by the flat radio spectrum (Fender et\,al.\ 1997), which we 
interpret as optically thick emission from jets oriented close to 
the line of sight.    

\begin{figure} 
\vspace*{0.2cm} 
\begin{center}
\begin{tabular}{ccc}
  \epsfxsize=5.5cm \epsfbox{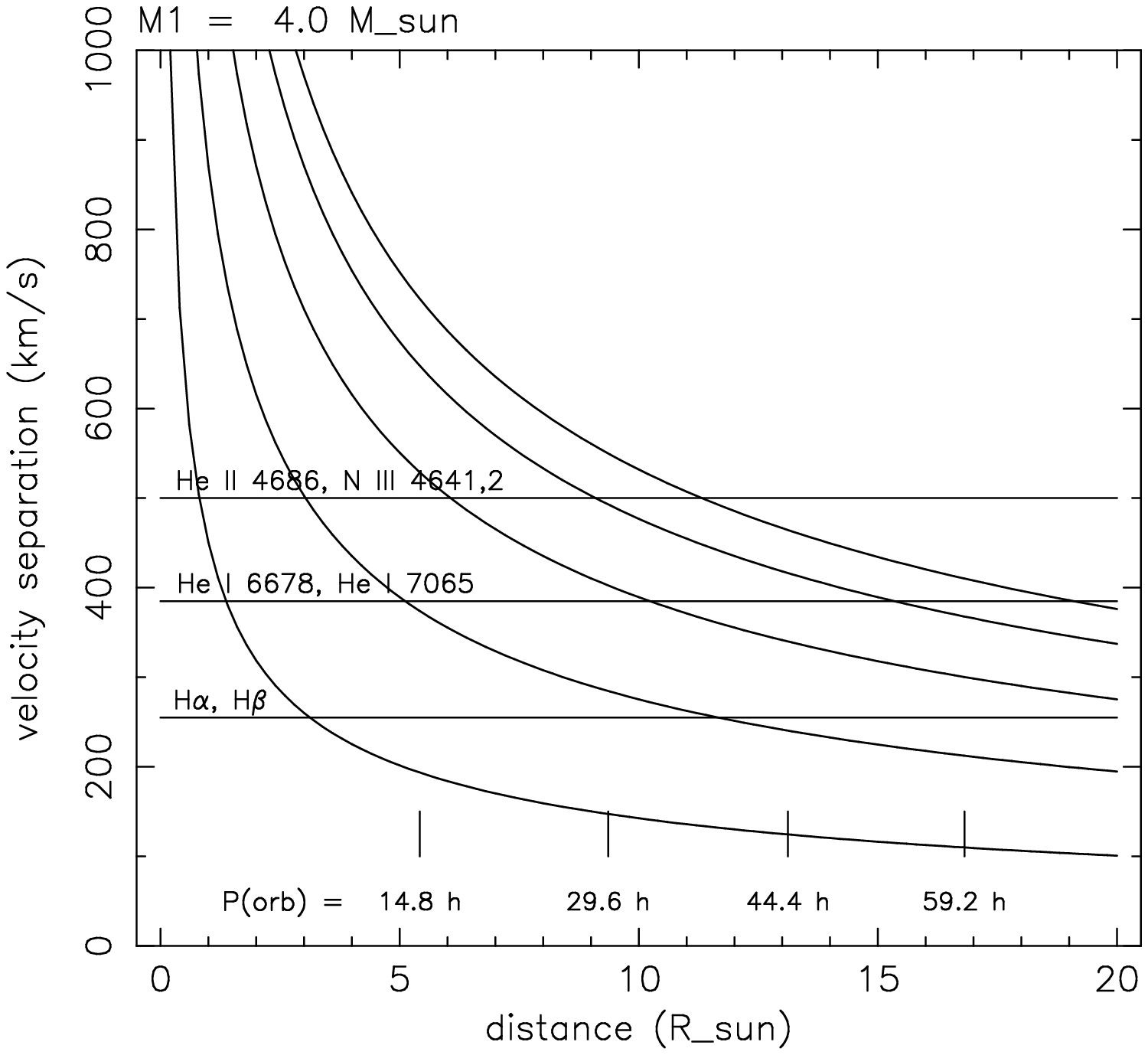} &
  \epsfxsize=5.5cm \epsfbox{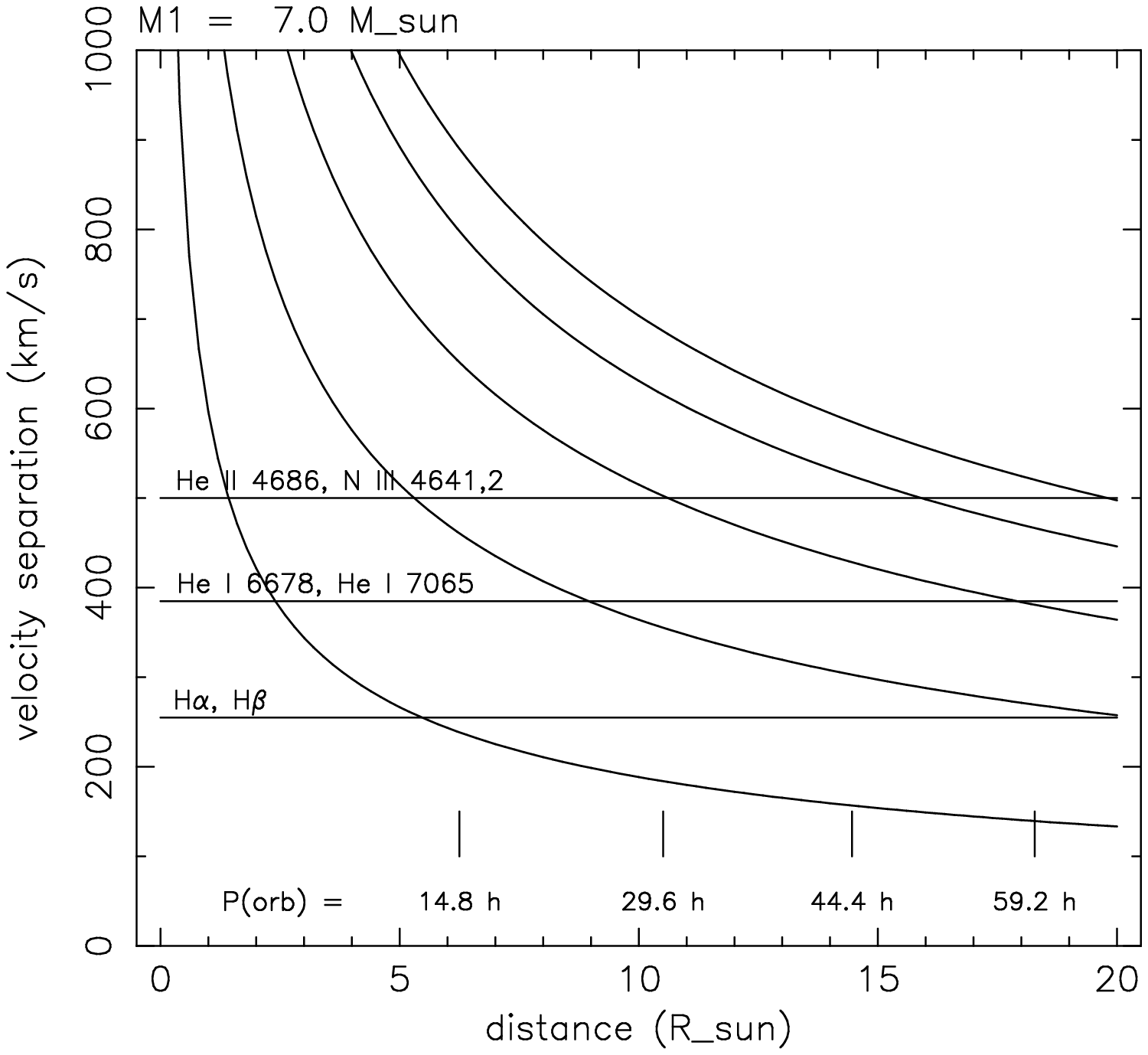} & 
  \epsfxsize=5.5cm \epsfbox{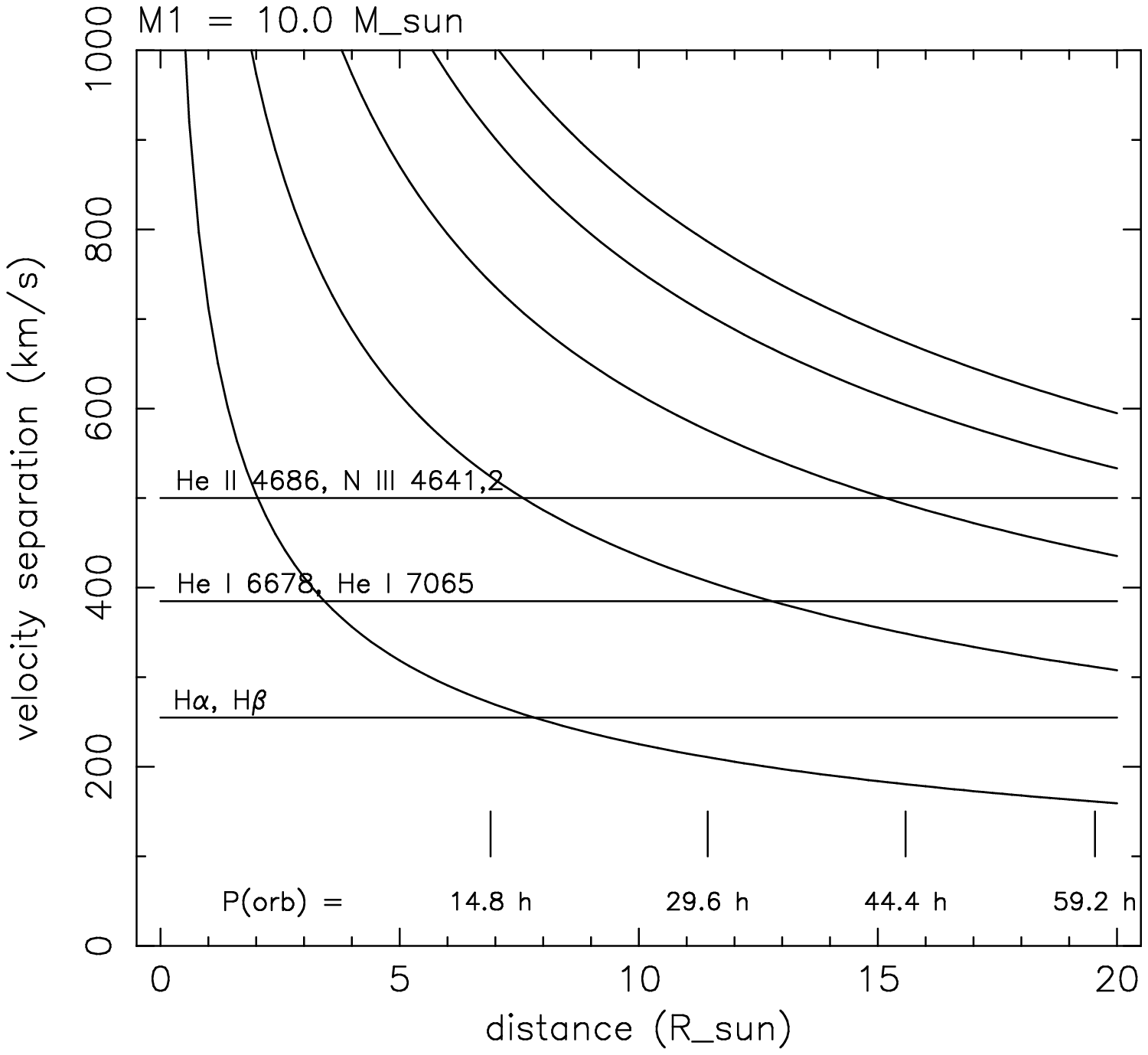}   
\end{tabular} 
\end{center}
\caption{The theoretical velocity separation of the line peaks as 
  a function of the distance from a compact star of mass equal to 
  4.0, 7.0 and 10.0~M$_\odot$ (from left to right), assuming a 
  geometrically thin, Keplerian accretion disk. 
  The curves correspond, from top to bottom, to the orbital 
  inclinations $i$ = 75, 60, 45, 30 and 15$^\circ$ respectively. 
  The horizontal lines are the velocity separations deduced from  
  our 1998 observations. The short vertical lines mark the orbital 
  separation of the system for multiples of the suspected 14.8-hr 
  orbital period. }
\end{figure}   
 
\section{Spectrum in the Low-hard state}    

\subsection{General features} 

Figure~3 shows the high-resolution (1.3~\AA) spectra that we obtained on 
1999 April 12. For comparison we also show the lower-resolution (3~\AA) 
summed spectrum from the 1997 May observations, when the system was in a 
previous low-hard state. In the 1999 spectra both the H$\alpha$ and the 
H$\beta$ lines are prominent,  
with equivalent widths (EWs) of $-8.5\pm 0.3$ and $-4.0 \pm 0.5$~\AA\ 
respectively. The H$\alpha$ line appears to be single-peaked, 
in contrast to the more clearly double-peaked profile observed 
in the high-soft state. For 
H$\beta$, the signal-to-noise of the spectrum does not allow us to 
determine the line profile. However, we do not see strong evidence of 
two peaks. 

The other prominent line is He\,{\scriptsize II} $\lambda$\,4686 
(EW = $-6.0 \pm 0.5$ \AA). It shows two clearly resolved peaks, unlike 
the Balmer lines. 
The N\,{\scriptsize III} $\lambda \lambda$\,4641,4642 lines are detected 
(with a limit to the EW $> -1.8$~\AA) but are much weaker than in 
1998 April and August (EW $\approx -(2.5 - 3.0)$~\AA) 
during the high-soft state. Other lines that can be identified are 
weak He\,{\scriptsize I} $\lambda$\,4922, and 
He\,{\scriptsize I} $\lambda$\,6678 emission. In the 1997 May spectrum, 
the He\,{\scriptsize I} $\lambda$\,6678 line is also found, and 
the N\,{\scriptsize II} $\lambda$\,6505 line is visible. 

\begin{figure}
\vspace*{0.35cm}
\begin{center}
\epsfxsize=12.75cm 
\epsfbox{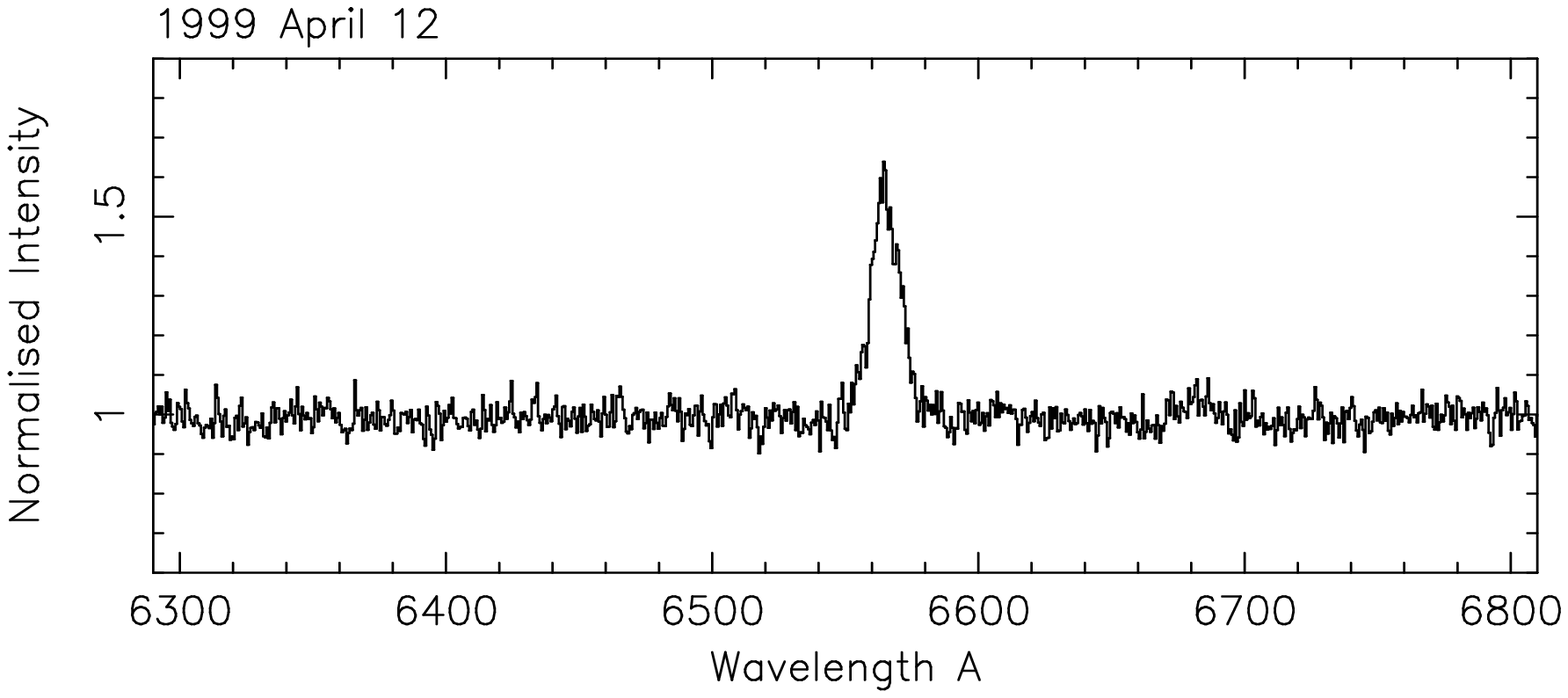}  
\end{center} 
\begin{center}
\epsfxsize=12.75cm 
\epsfbox{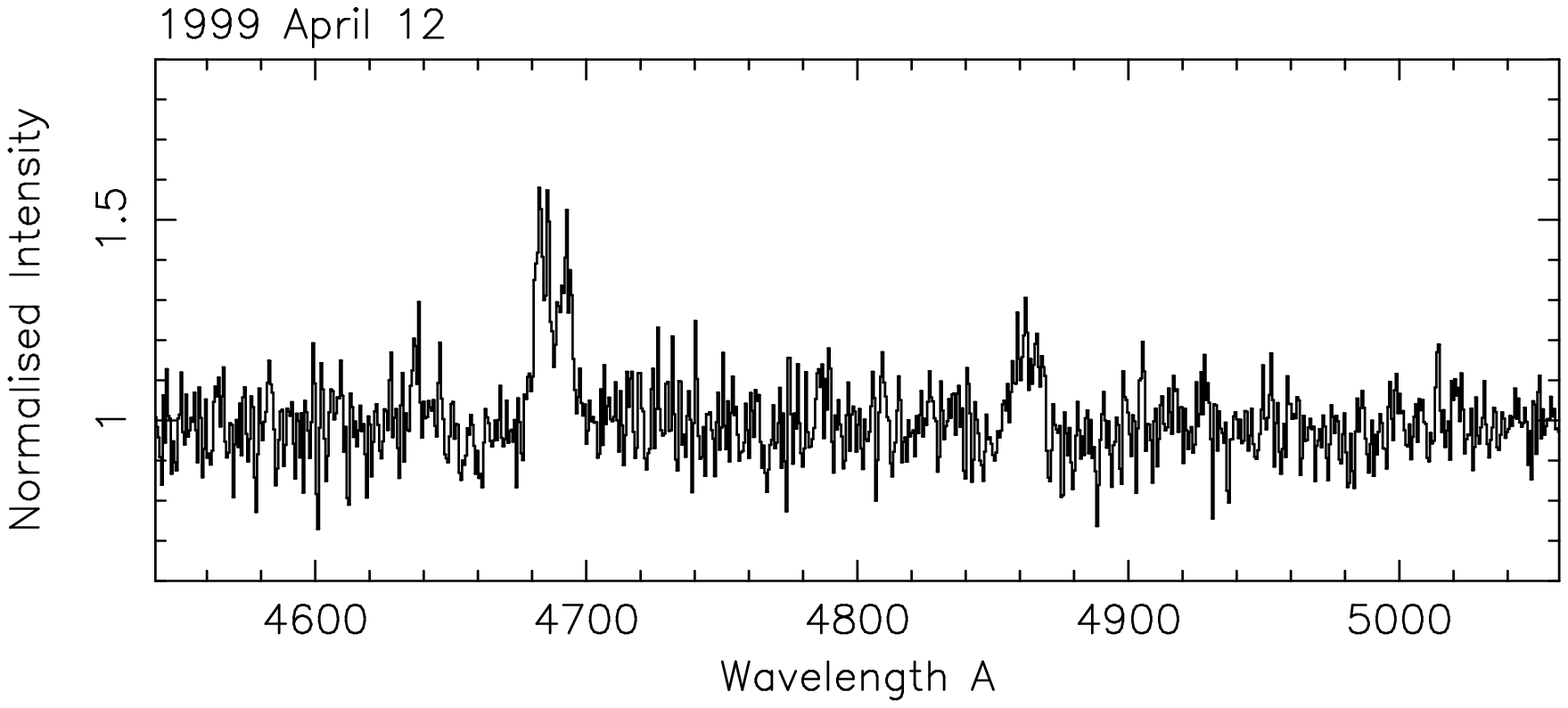}    
\end{center} 
\begin{center}
\epsfxsize=12.75cm 
\epsfbox{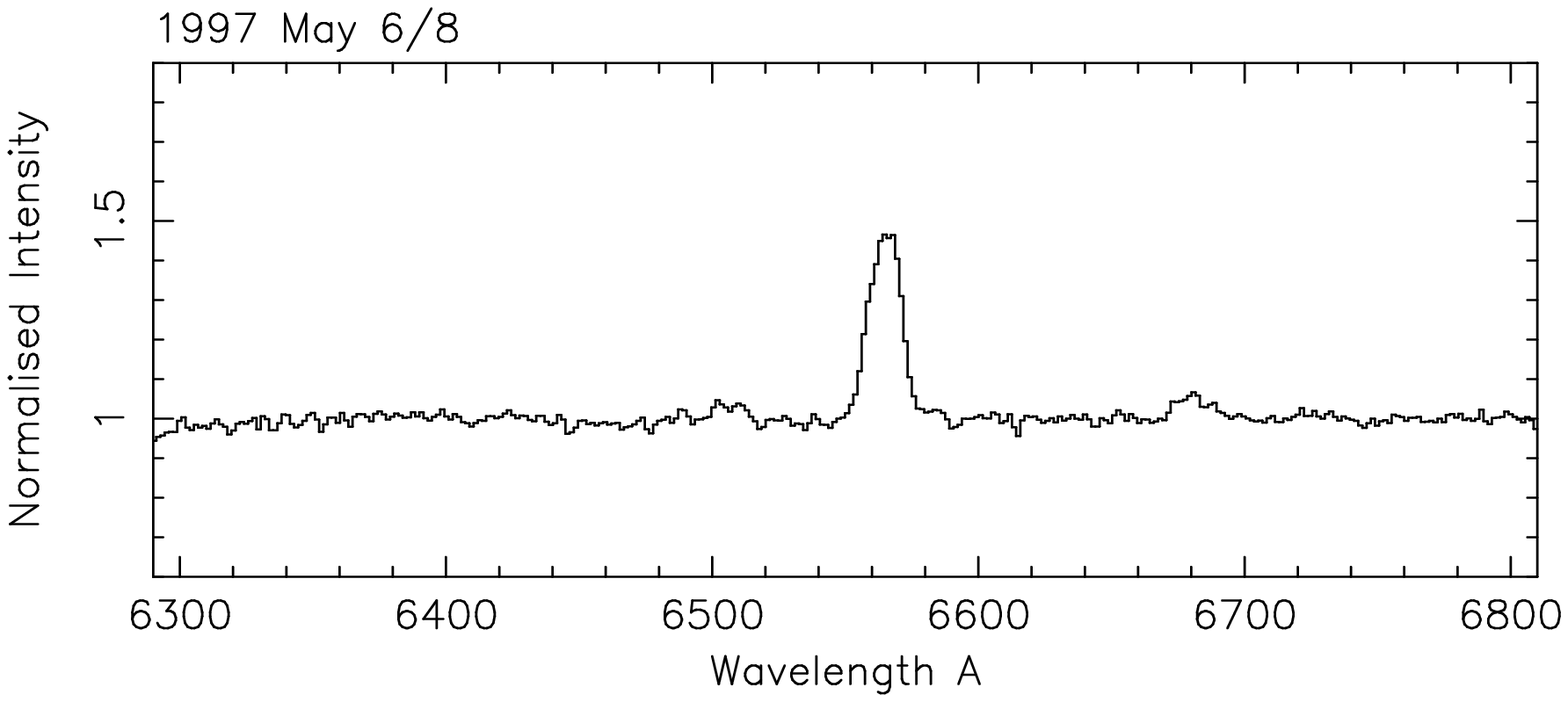}    
\end{center} 
\caption{The high-resolution (1.3~\AA) spectra that we obtained on 1999 
  April 12 when the system was in a low-hard state. About five days 
  later the system entered an X-ray quiescent state. The spectra are 
  centred at H$\alpha$ (top panel) and at H$\beta$/He\,{\scriptsize II} 
  (middle panel). The lower-resolution (3~\AA) summed spectra, centred 
  at H$\alpha$, obtained in 1997 May (low-hard state), is also shown 
  for comparison (bottom panel).}
\end{figure} 

\subsection{Are the emission lines single-peaked or double-peaked during 
  the low-hard state?}   

Medium-resolution ($\approx 2.5$~\AA) spectra obtained by 
Smith, Filippenko \& Leonard (1999) 
during the low-hard state in 1996 May showed a broad, flat-topped H$\alpha$ 
line. The profile resembles a line with two unresolved peaks, and  
Smith, Filippenko \& Leonard (1999) interpreted the H$\alpha$ line as 
being double-peaked. However, our 1997 May observations, when \gx339 was 
still in its low-hard state, did not show a double-peaked H$\alpha$ line.  
Because of the low resolution (3 \AA) of our spectra, we were unable to 
verify whether the H$\alpha$ line was single- or doubled-peaked.   
The higher-resolution (1.3 \AA) spectrum that we 
obtained on 1999 April 12 allows us to carry out a more quantitative 
analysis. 

The profiles of the H$\alpha$ and the He\,{\scriptsize II} $
\lambda$\,4686 lines in the high-soft and low-hard states are shown 
in Figure 4. The H$\alpha$ profiles 
of \gro1655 obtained during its high-soft state in 1996 June and the 
hard X-ray outburst in 1994 September are also shown for 
comparison.

We have used the normalised difference spectra to examine the 
differences in the flux distribution in the line profiles from 
each observing epoch. For each spectrum, we removed   
the continuum by fitting and subtracting a low-order polynomial 
in the velocity range between $-4000$~\kms\ and $+4000$~\kms. The 
residual line flux was then normalised to unity. Finally, the line 
spectrum obtained in the low-hard state was subtracted from the 
spectrum obtained in the 1998 high-soft state.  

In the left panel of Figure 5, we show the difference 
profile of the H$\alpha$ lines for the 1998/1999 observations. The 
two peaks in the 1998 observations are clearly seen above the noise 
level. The peaks are also visible, though less prominently, in 
the H$\alpha$ difference profile for the 1997/1998 observations 
(Fig.~5, middle panel). This suggests that the H$\alpha$ profiles are  
different in the high-soft and the low-hard states. While the 
high-soft state is characterised by a double-peaked profile, the 
line probably has a single-peaked profile in the low-hard state. 
 
The He\,{\scriptsize II} $\lambda$\,4686 line is double-peaked in  
both our 1998 and 1999 spectra, making the difference profile 
(Fig.~5, right panel) less straightforward to interpret. The overall 
shape results from the difference between the widths of the lines.  

Our observations show that both the H$\alpha$ and the 
He\,{\scriptsize II} $\lambda$\,4686 lines are double-peaked in the 
high-soft state. In the low-hard state, 
the He\,{\scriptsize II} $\lambda$\,4686 line clearly shows two 
peaks, but the profile of the H$\alpha$ line is better described as 
single-peaked. This indicates (i) different emission regions for 
H$\alpha$ in the high-soft and the low-hard states, (ii) different 
line-formation mechanisms in the two states, and (iii) different 
response of the outer accretion disk under irradiation by soft and 
hard X-rays. These issues  will be discussed in more detail in the 
later sections.   

\begin{figure} 
\vspace*{0.25cm}  
\begin{center}
\begin{tabular}{cc}
  \epsfxsize=6.2cm \epsfbox{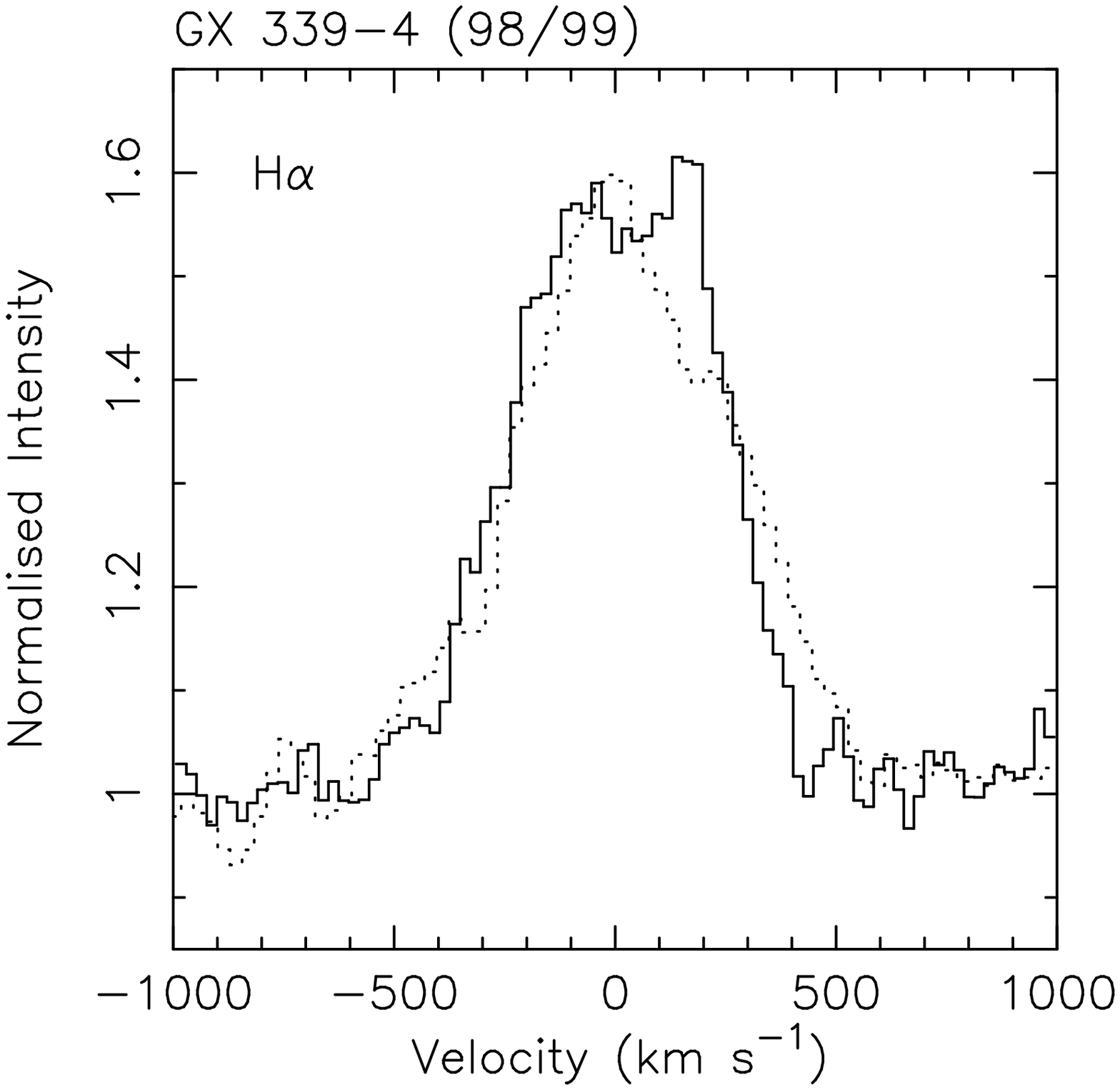} & 
  \epsfxsize=6.2cm \epsfbox{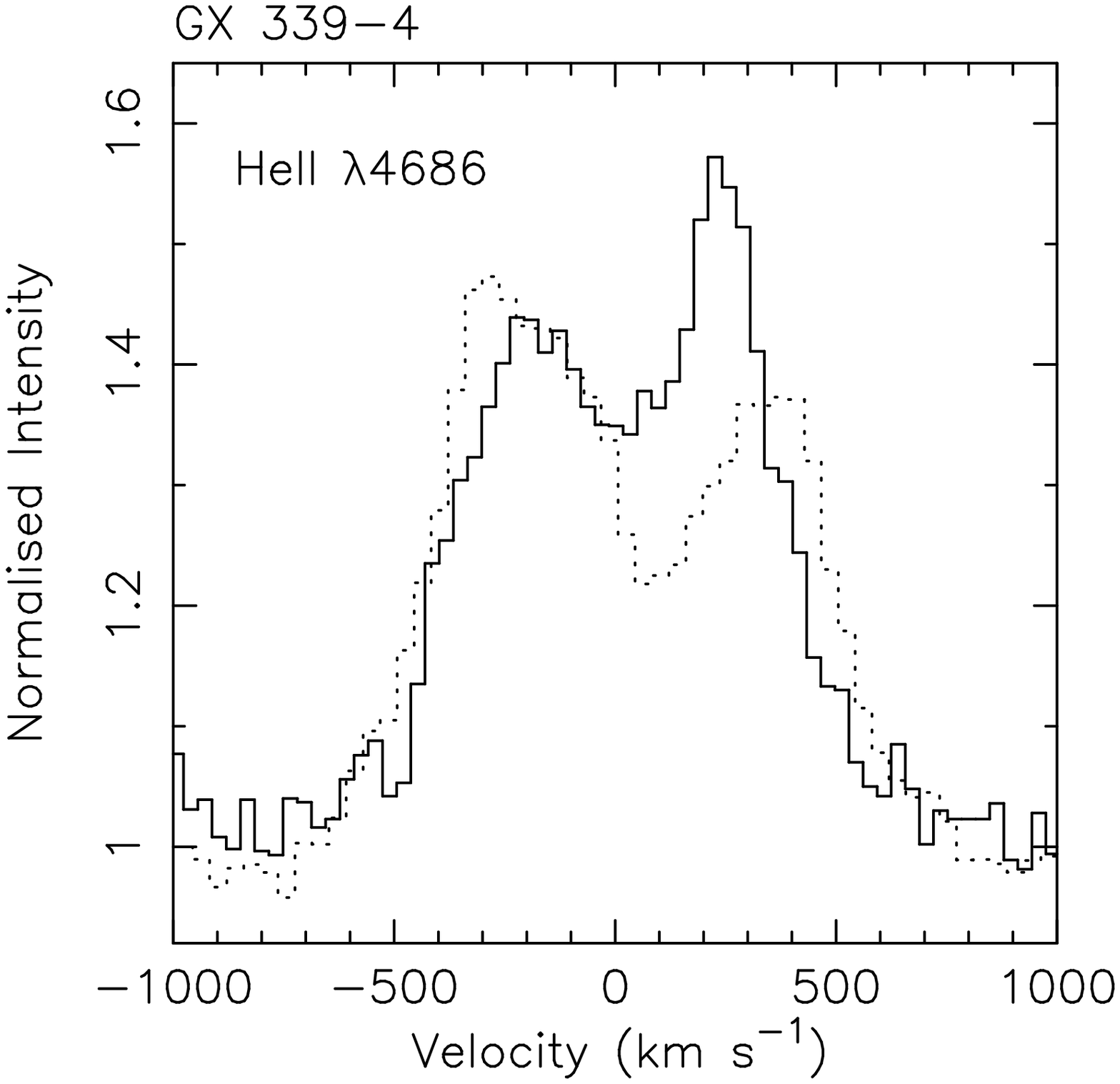} \\ 
  \epsfxsize=6.2cm \epsfbox{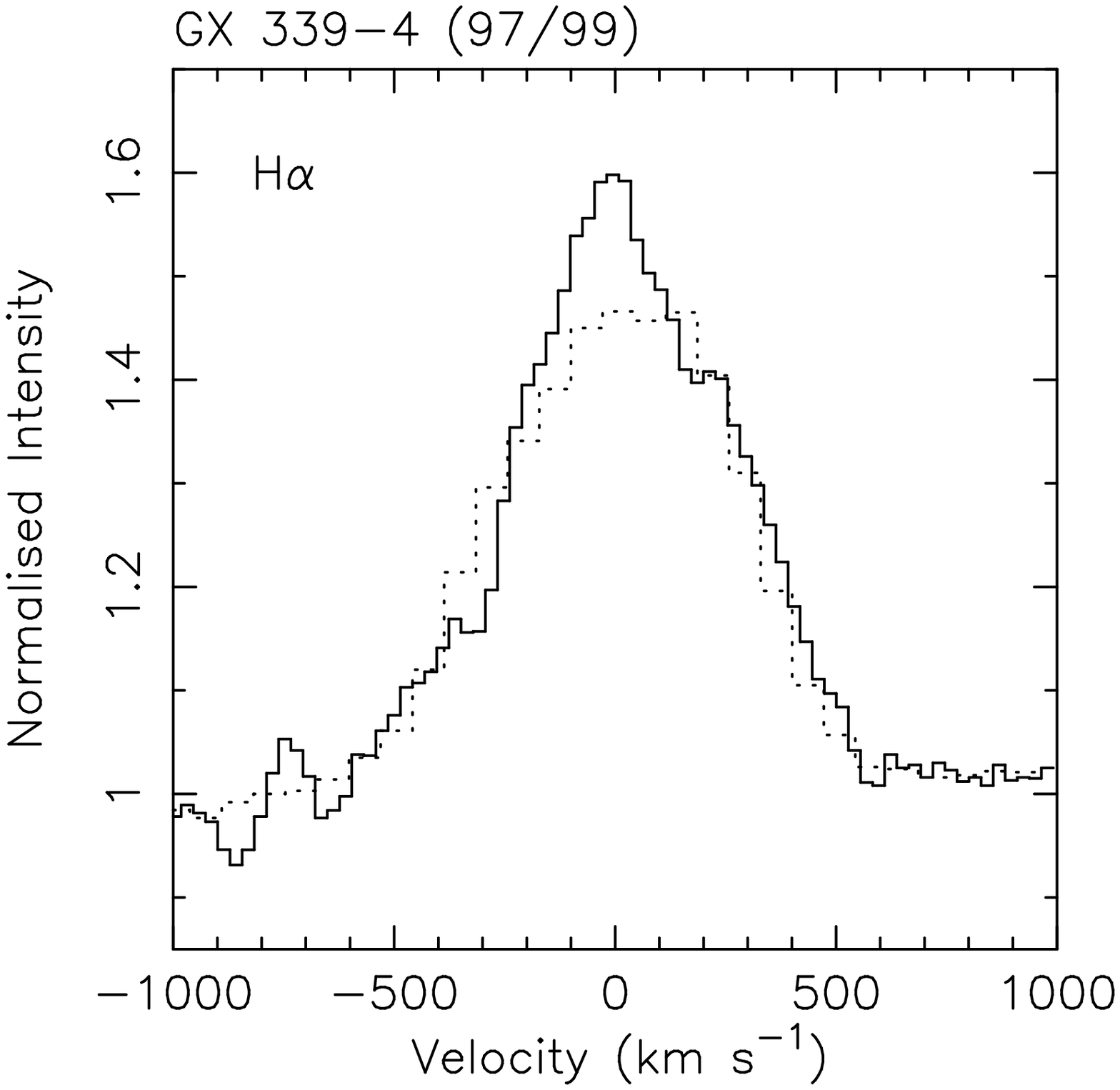} & 
  \epsfxsize=6.2cm \epsfbox{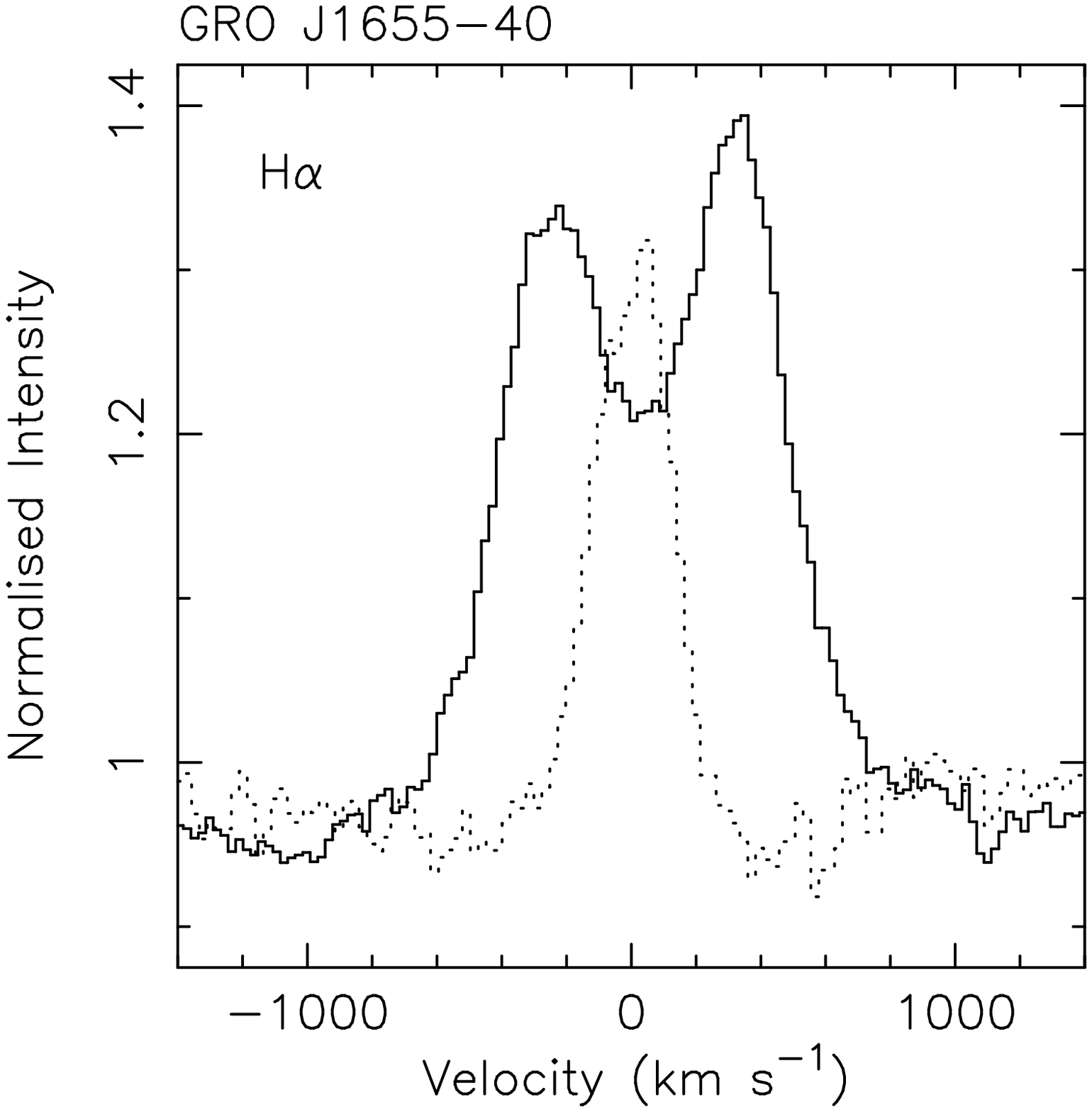} 
\end{tabular} 
\end{center}
\caption{Profiles of  H$\alpha$ (top-left panel) and 
 He\,{\scriptsize II} $\lambda$\,4686 (top-right panel) 
 during the high-soft state in 1998 April and the low-hard state in 
 1999 April (represented by solid and dotted lines respectively). 
 We smoothed the 1999 low-hard state profiles by averaging over three 
 data points. We show the comparison of the H$\alpha$-line profiles 
 for the 1997 and 1999 low-hard states (dotted and solid lines 
 respectively, bottom-left panel). The 1997 data are not smoothed. 
 Profiles of the H$\alpha$ line from the high-orbital-inclination 
 ($i \approx 70^\circ$) BHC \gro1655 are shown (bottom-right panel). 
 The broad, double-peaked H$\alpha$ profile of \gro1655 (solid line) 
 was obtained on 1996 June 10 when the system was in a 
 high-soft state; the narrow H$\alpha$ profile (dotted line) was 
 obtained on 1994 September 3 when it was active in the hard 
 X-ray bands. The normalisations of the continua in the \gro1655 
 spectra fall below unity because of the broad 
 H$\alpha$ absorption component (Soria, Wu \& Hunstead 2000). }
\end{figure}   

\subsection{Comments on the line profiles and line widths} 
 
\subsubsection {He\,{\scriptsize II} $\lambda$\,4686 and 
N\,{\scriptsize III} $\lambda \lambda$\,4641,4642}  

In spite of the similar double-peaked profiles of 
He\,{\scriptsize II} $\lambda$\,4686 
in the two X-ray spectral states, the peak separations  
are clearly different: 490$\pm$66 \kms\ for the high-soft state  
and 600$\pm$50 \kms\ for the low-hard state (Fig.~4, top-right panel). 
The line widths are also different in the two states 
(Fig.~5, right panel). The larger separation 
implies that the emission region is at a smaller radial distance from 
the compact object. The larger peak separation in the low-hard 
state may be due to the fact that the soft X-ray component 
was weak and unable to set up the temperature-inversion layer 
(see \S 5) in the outer disk regions.  

We note that the N\,{\scriptsize III} $\lambda \lambda$\,4641,4642 
lines in 1999 April (Fig.~3) were much weaker than in 1998 
(see Fig.~3 in Paper I). 
The N\,{\scriptsize III} $\lambda \lambda$\,4641,4642 lines  
are due to Bowen fluorescence, which is pumped by the 
O\,{\scriptsize III} $\lambda$\,374 multiplet resulting  
from the cascade O\,{\scriptsize III} 
2p3d~($^3$P$^0_2$) $\rightarrow$ 2p3p~($^3$S,$^3$P,$^3$D) 
 $\rightarrow$ 2p3s~($^3$P$_0$)  $\rightarrow$ 2p$^2$~($^3$P$_2$) 
transitions primarily pumped by the 
UV He\,{\scriptsize II} $\lambda$\,303.78 (Lyman-$\alpha$) line 
(see e.g.\ Deguchi 1985; Schachter, Filippenko \& Kahn 1989).
The He\,{\scriptsize II} Lyman-$\alpha$ line can be 
produced by the photoionisation-recombination process, and is 
therefore associated with the He\,{\scriptsize II} $\lambda$\,4686 
line. Hence, one might expect to observe strong 
N\,{\scriptsize III} $\lambda \lambda$\,4641,4642 emission when 
He\,{\scriptsize II} $\lambda$\,4686 is prominent. This is what we 
have observed in \gx339 and other BHCs (e.g.\ \gro1655 and 
RXTE~J1550$-$564) during the high-soft states. The situation 
is different during the low-hard state of \gx339 in 1999 April: 
the N\,{\scriptsize III} $\lambda \lambda$\,4641,4642 lines are weak, 
yet the He\,{\scriptsize II} $\lambda$\,4686 remains prominent. 
A similar phenomenon was also observed in the BHC GRO~J0422$+$32 during 
a hard X-ray outburst (Casares et al.\ 1995). 

It was suggested that 
the absence of N\,{\scriptsize III} $\lambda \lambda$\,4641,4642 lines 
in GRO~J0422$+$32 was due to insufficient X-ray irradiation 
(Casares et al.\ 1995). One would therefore  
expect {\it both} the He\,{\scriptsize II} $\lambda$\,4686 and the 
N\,{\scriptsize III} $\lambda \lambda$\,4641,4642 lines to be weak. 
Here (and in various other systems), our observations show 
a strong He\,{\scriptsize II} $\lambda$\,4686 line both in the 
high-soft and low-hard states, but significantly weaker 
N\,{\scriptsize III} $\lambda \lambda$\,4641,4642 lines in the 
low-hard state.  Clearly, 
there must also be other factors. We propose that the 
weakening of the N\,{\scriptsize III} $\lambda \lambda$\,4641,4642 
emission from BHCs in the low-hard state is due to optical depth 
effects. When the He\,{\scriptsize II} region 
becomes optically thick to UV continuum, the 
He\,{\scriptsize II} Lyman-$\alpha$ and   
O\,{\scriptsize III} $\lambda$\,374 photons are quenched. If  
it is still relatively transparent to the optical continuum,   
He\,{\scriptsize II} $\lambda$\,4686 photons can escape. 
Without efficient pumping by He\,{\scriptsize II} Lyman-$\alpha$  
and O\,{\scriptsize III} $\lambda$\,374 
photons, the N\,{\scriptsize III} $\lambda \lambda$\,4641,4642 lines 
are suppressed. If this is the situation in \gx339, the 
photoionisation-recombination processes which produce the 
He\,{\scriptsize II} $\lambda$\,4686 lines may be different 
during the high-soft and low-hard X-ray states (cf.\ Case A, Case B 
and Case C recombination processes, see Osterbrock 1989;  
Xu et al.\ 1992).  

\begin{figure} 
\vspace*{0.25cm}  
\begin{center}
\begin{tabular}{ccc}
  \epsfxsize=5.5cm \epsfbox{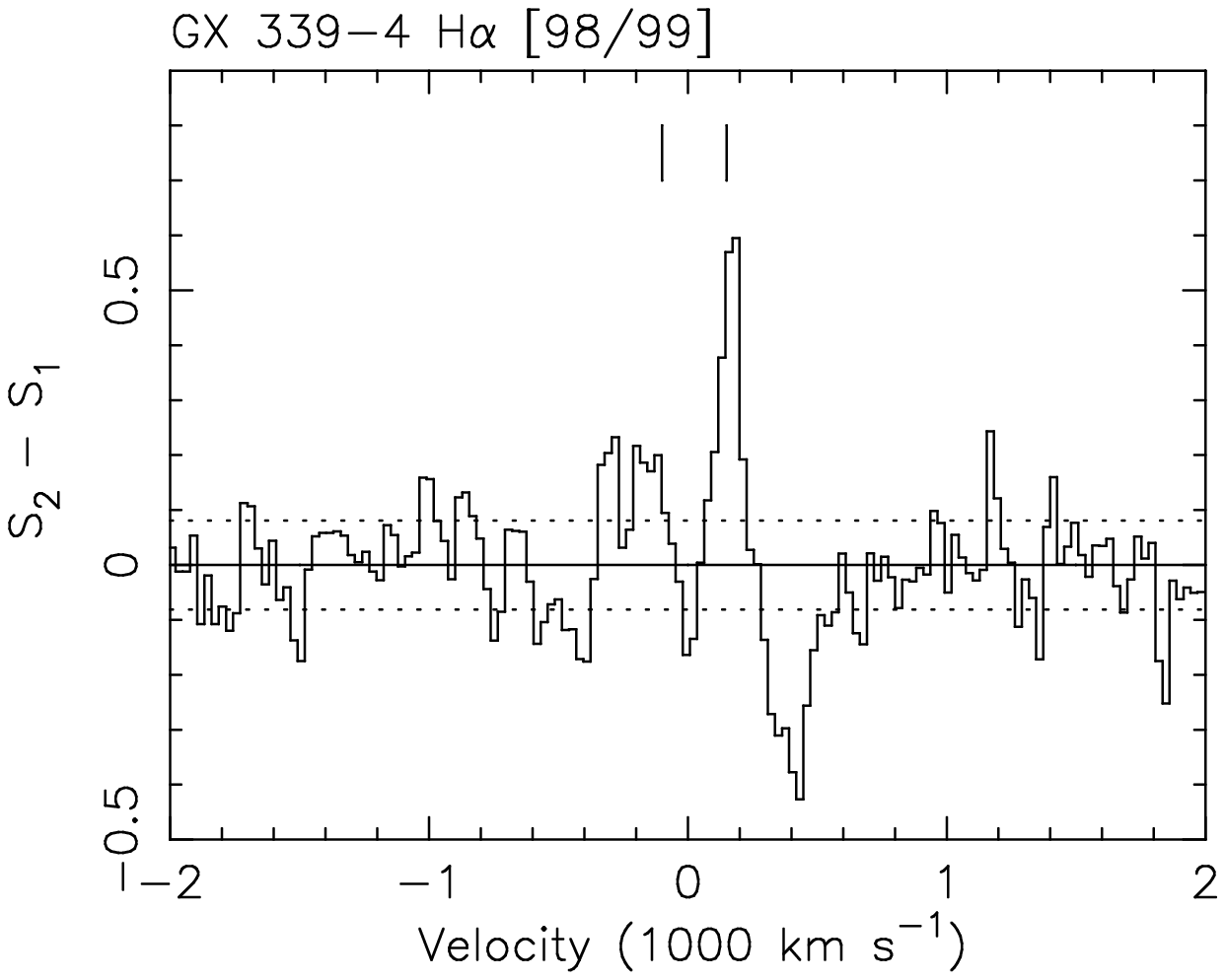} & 
  \epsfxsize=5.5cm \epsfbox{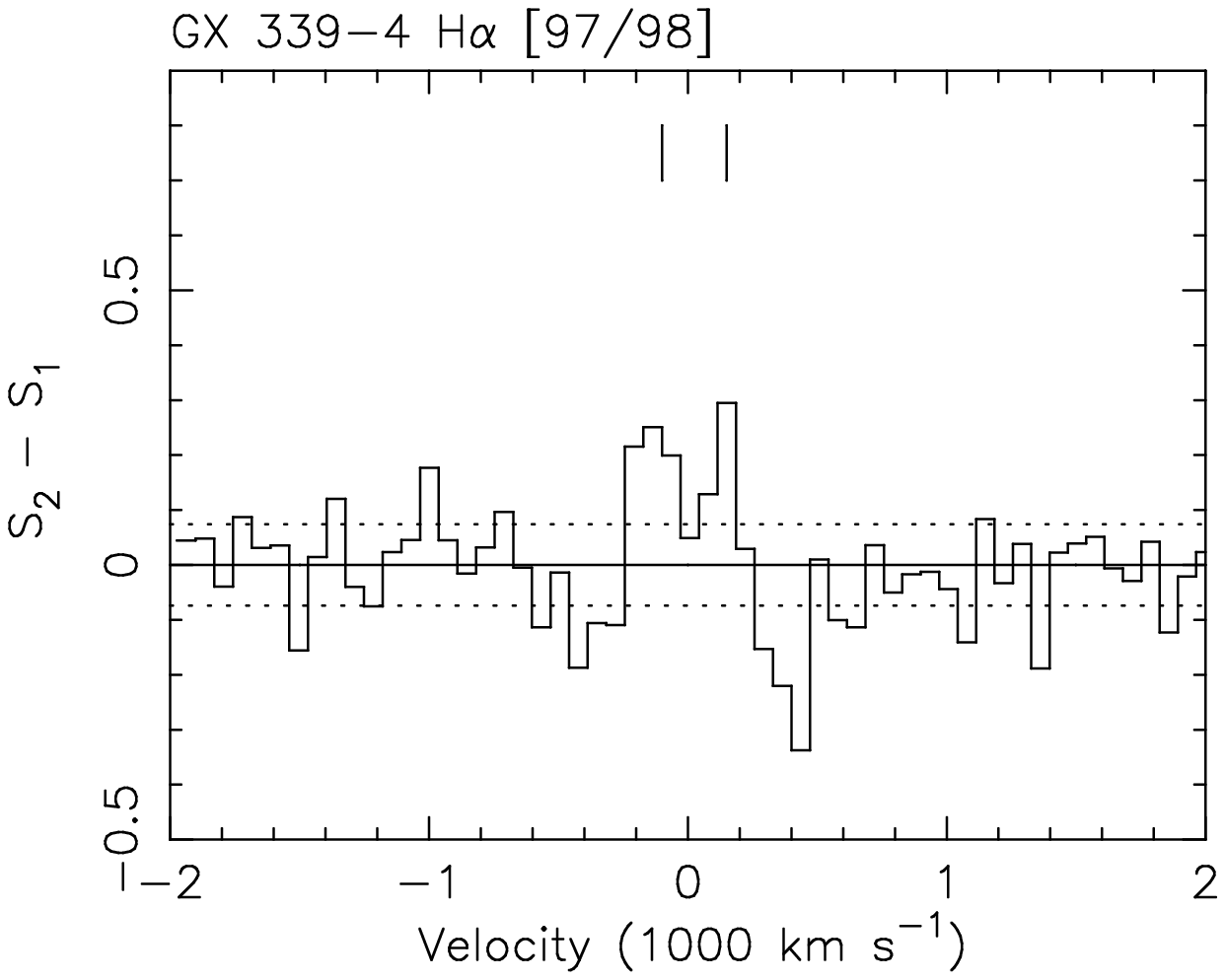} & 
  \epsfxsize=5.5cm \epsfbox{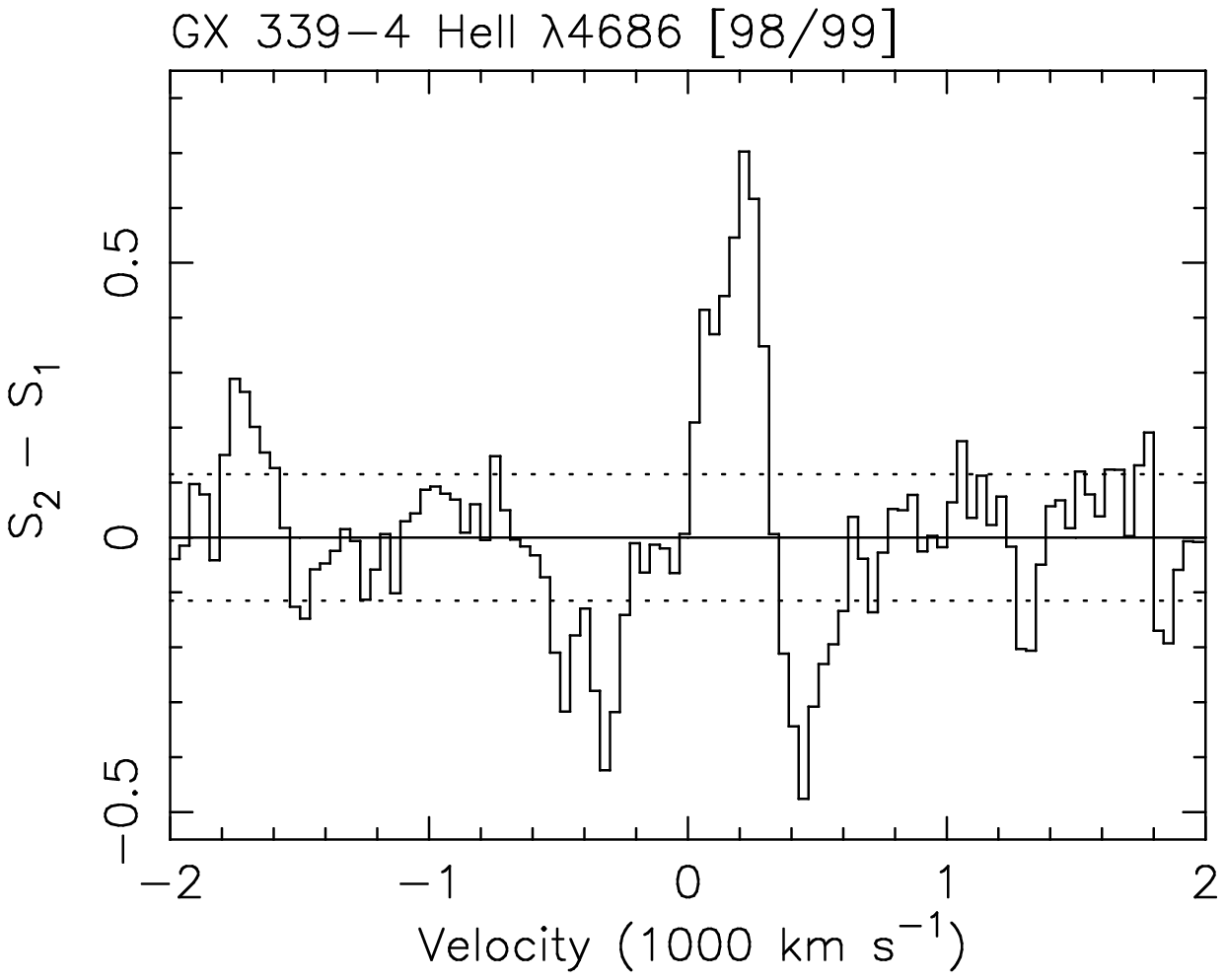} \end{tabular} 
\end{center}
\caption{Normalised difference H$\alpha$ line profiles  
 of the 1998/1999 and 1997/1998 observations (left and middle panels 
 respectively). The locations of the peaks of the line obtained in the 
 1998 observations are marked (vertical lines).  
 We calculated the difference profiles by subtracting 
 the profile obtained in the 1997/1999 low-hard states (S$_1$)
 from those obtained in the 1998 high-soft state (S$_2$). 
 The integrated line fluxes (after 
 continuum removal) are normalised to unity. The 1-$\sigma$ noise level  
 (dotted lines) is calculated from the continuum in the velocity 
 ranges between $-4000$ \kms\ and $-1000$ \kms, and 
 between 1000 \kms and 4000 \kms. The difference profile of 
 He\,{\scriptsize II} $\lambda$\,4686 for the 1998/1999  observations 
 is also shown (right panel). Note that the excess in the flux at 
 velocities $< -1500$~\kms\ is due to the 
 N\,{\scriptsize III} $\lambda \lambda$\,4641,4642 lines, which were  
 more prominent in 1998 than in 1999. }
\end{figure}   
   
\subsubsection{Line width of the H$\alpha$ emission line}

The profiles of the H$\alpha$ and the
He\,{\scriptsize II} $\lambda$4686 lines observed from \gx339 in the high-soft 
and low-hard states are shown in the first three panels of Figure 4. 
The width of H$\alpha$ is similar in the two states, in contrast to 
He\,{\scriptsize II} $\lambda$4686, in which case the line width increases 
with the hardness of the X-rays. 

We interpret the double-peaked lines in the high-soft state 
of \gx339 (and in the high-soft state of other BHCs, like \gro1655)
as originating from the temperature-inversion layer created by soft X-ray 
irradiation on the accretion disk surface. 
The single-peaked (flat-topped or round-topped)
Balmer lines observed in the low-hard states of \gx339 
are probably emitted from an outflow, possibly a dense wind from the 
evaporating atmosphere of the accretion disk.
Theoretical studies (Murray et\,al.\ 1995; 
Murray \& Chiang 1998) have shown that
if the outflow/wind has a substantial 
velocity component parallel to the disk plane, single-peaked lines with 
widths similar to the Keplerian disk velocities can be produced. 
The profiles will appear to be round-topped 
or flat-topped, depending on the outflow velocity and density 
profiles (see Chapter 14 in Mihalas 1978), as in the case of emission 
lines from windy massive stars.
 
Single-peaked Balmer emission lines were also observed from \gro1655 
in 1994 August -- September, in between strong hard X-ray flares. 
(Soria, Wu \& Hunstead 2000).
On that occasion, the lines were much narrower than the double-peaked 
lines observed during the high-soft state of the system 
(Fig.~4, bottom-right panel). 
The narrow, single-peaked lines from \gro1655 are probably emission 
from an extended, optically thin cocoon 
above the disk plane, with lower rotational velocity than the disk 
or the disk wind. In the high-inclination system \gro1655 
($i \approx 70^{\circ}$), most of this extended, optically-thin envelope 
was seen projected onto the sky and was therefore a source of emission lines. 
Narrow emission lines have not been observed in 
the low-inclination system \gx339, where the accretion disk is seen almost 
face-on and any thin cocoon would be seen projected onto the disk surface.

\section{Line emission region}   

\subsection{A plane-parallel model} 

We consider a simple plane-parallel model (Fig.~6) to illustrate that 
a temperature-inversion layer can be set up at the surface of an 
accretion disk under strong soft X-ray illumination but not under hard 
X-ray illumination. The model is a modification of the model constructed 
by Milne (1926) for the illumination of a star by its companion in a 
close binary system.  
In Milne's original formulation the incident radiation is at optical 
wavelengths, and the cooling is via emission of optical radiation. In 
BHCs, the incident radiation on the accretion disk consists of soft and 
hard X-rays emitted from regions close to the central compact object. 
We therefore need to take this difference into account. 

In our model, we consider that the incident radiation has 
soft and hard X-ray components, for which the opacities are 
respectively larger and smaller than for the optical radiation.  
This modification is essential in the study of accretion 
disks irradiatively heated by X-rays, as soft X-rays are easily 
absorbed by neutral and weakly ionised 
matter (via bound-free transitions) at the disk surface, while  
hard X-rays are attenuated only at large depths  
where the matter density is sufficiently high. The other assumptions 
are the same as those in the Milne (1926) model, namely 
(i) the emission region has a plane-parallel geometry, (ii) the 
system is in radiative equilibrium, and (iii) the principle of 
superposition is applicable.   

\begin{figure} 
\vspace*{0.5cm}  
\begin{center}
  \epsfxsize=6.cm 
  \epsfbox{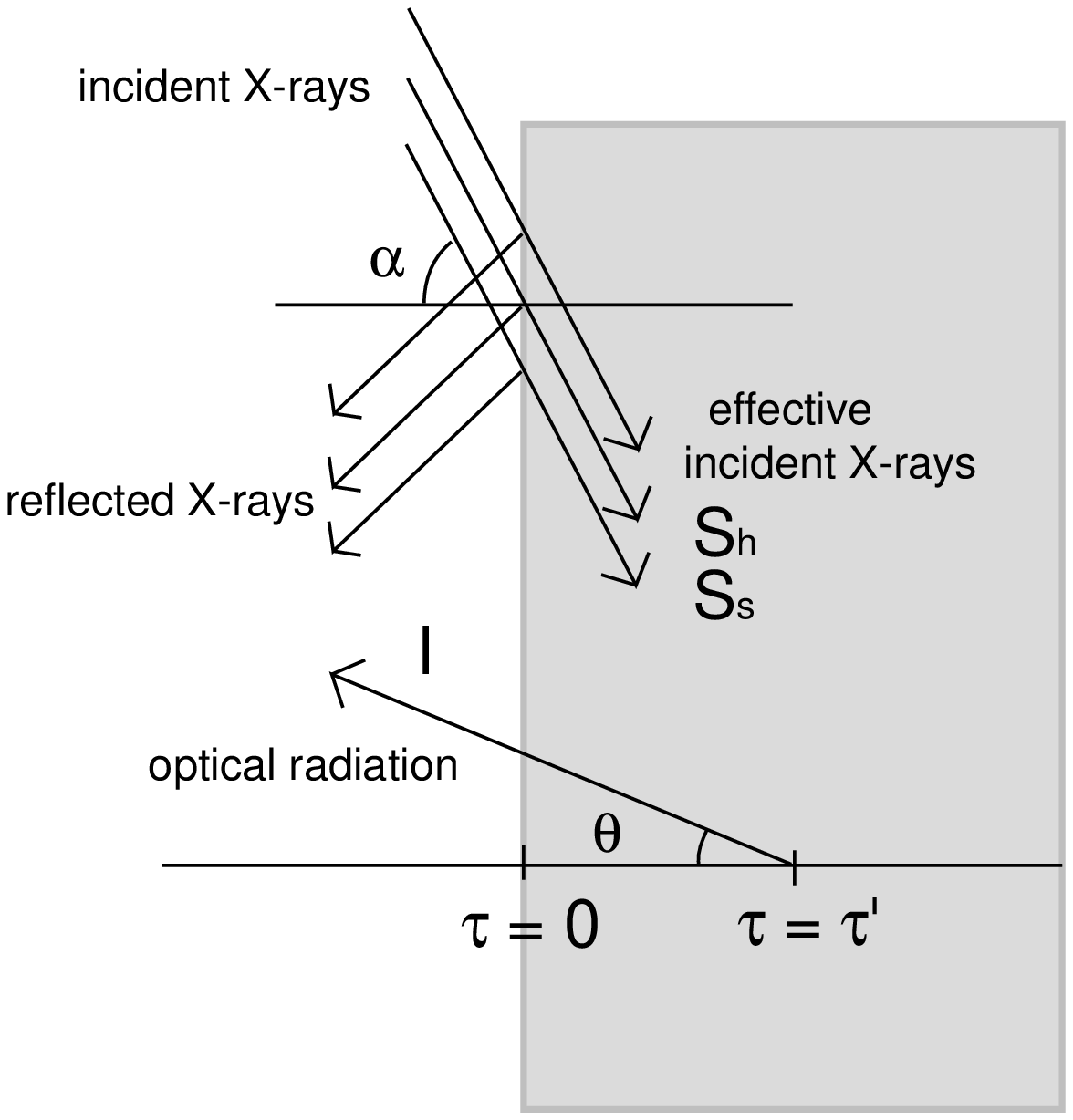}  
\end{center}
\caption{The geometry of the modified Milne (1921) plane-parallel model.}   
\end{figure}    

As the radiative-transfer equations are linear, the local 
temperature stratification is given by  
\begin{eqnarray} 
 T(\tau) & = & \biggl[ {\pi \over \sigma} 
   \biggl( B_{\rm x}(\tau) + B_{\rm d}(\tau) \biggr) \biggr]^{1/4}   
   \nonumber \\ 
   & \equiv & \biggl[ {\pi \over \sigma} B(\tau) \biggr]^{1/4}
\end{eqnarray}  
(see Milne 1926), where $B_{\rm x}(\tau)$ is the component of the 
radiation due to irradiative heating,  
$B_{\rm d}(\tau)$ is the component due to viscous heating in the disk 
in the absence of X-ray irradiation, $\tau$ is the optical depth, and 
$\sigma$ is the Stefan-Boltzmann constant. We assume that the disk is 
cooled by emission of optical radiation, and therefore, $\tau$ refers 
to the optical depth in the optical band, along the direction normal 
to the disk surface. The disk component is given by 
\begin{equation}   
   B_{\rm d}(\tau)\ =\ {3  \over 4}S_{\rm d}~\biggl[ 
    \tau \biggl( 1 - {\tau \over {2 \tau_{\rm tot}}} \biggr) 
    + {2 \over 3} \biggr]   
\end{equation}   
(e.g.\ Dubus et al.\ 1999), where the viscous energy flux of the disk 
$S_{\rm d} = \sigma  T_{\rm eff}^4 / \pi$ and $\tau_{\rm tot}$ is 
the total opacity of the accretion disk in the vertical direction.  
The effective temperature $T_{\rm eff}$, which is determined  
by balancing the energy generated by viscous heating and by 
radiative loss (see e.g.\ Smak 1984), is given roughly by 
\begin{equation} 
  \sigma T_{\rm eff}^4 \  \approx \  
  {9 \over 8 } \nu \Sigma \Omega_{\rm K}^2  \ ,  
\end{equation}  
where $\nu$ is the viscosity coefficient, $\Sigma$ is the surface 
density of the accretion disk, and $\Omega_{\rm K}$ is the Keplerian 
angular frequency.  
 
Suppose the incident radiation consists of parallel beams of soft and 
hard X-rays, with effective fluxes $\pi S_{\rm s}$ and $\pi S_{\rm h}$ 
per unit area normal to the beams, and making an angle 
$\alpha$ with the normal to the disk plane. The beams are absorbed 
exponentially. We denote the absorption coefficient of the soft X-rays 
by $k_{\rm s}\kappa$, and that of the hard X-rays by $k_{\rm h}\kappa$, 
where $\kappa$ is the absorption coefficient of the optical radiation. 

At the optical depth $\tau$, the soft and hard 
X-ray fluxes are attenuated to 
$\pi S_{\rm s} e^{-k_{\rm s} \tau \sec \alpha} \cos \alpha$ and  
$\pi S_{\rm h} e^{-k_{\rm h} \tau \sec \alpha} \cos \alpha$ 
respectively.   
The absorbed flux of X-rays in depth $d\tau$ is therefore 
 $\pi (k_{\rm s} S_{\rm s} e^{-k_{\rm s} \tau \sec \alpha} +  
   k_{\rm h}S_{\rm h} e^{-k_{\rm h} \tau \sec \alpha})~ d\tau$.  
Radiative equilibrium requires 
\begin{equation} 
 \pi \big( k_{\rm s}S_{\rm s} e^{-k_{\rm s} \tau \sec \alpha} +  
   k_{\rm h}S_{\rm h} e^{-k_{\rm h} \tau \sec \alpha}\big) + \int_{4 \pi} 
   d\Omega~ I(\tau,\mu)\ 
   =\ 4\pi B_{\rm x}(\tau) \ ,  
\end{equation}  
where $I(\tau,\mu)$ is the intensity of the optical radiation 
from the irradiatively heated disk at 
the depth $\tau$ in the direction $\theta~ (= \cos^{-1}\mu)$.  

Recall the radiative-transfer equation 
\begin{equation} 
 \mu {d \over {d \tau}} I(\tau,\mu) \ 
  = \ I(\tau,\mu) - B_{\rm x}(\tau) \ . 
\end{equation}  
Multiplying the above equation by a differential solid angle $d\Omega$, 
integrating with respect to $\tau$ and using the condition of radiative 
equilibrium, we have 
\begin{equation} 
 \int_{4 \pi} 
   d\Omega~ \mu~I(\tau,\mu)  = \ 
 \pi \cos \alpha ~\big(
  {S_{\rm s}} ~ e^{-k_{\rm s} \tau \sec \alpha} +  
  {S_{\rm h}} ~ e^{-k_{\rm h} \tau \sec \alpha}\big) \ . 
\end{equation}  

With appropriate boundary conditions, the radiative-transfer and 
radiative-equilibrium equations can be solved by the method of 
successive approximations (Milne 1921, 1926). In the limit of a 
semi-infinite plane the first-approximation solution is 
\begin{eqnarray} 
 B_{\rm x}(\tau) & = & 
  k_{\rm s}{S_{\rm s}}~\bigg\{ 
   \biggl( {{\cos \alpha} \over {k_{\rm s}}}  + {1 \over 2}  
   \biggr) 
 \biggl[ \biggl({{\cos \alpha} \over {k_{\rm s}}} \biggr) - \biggl(
   {{\cos \alpha} \over {k_{\rm s}}}  - {1 \over 2}  
   \biggr)~e^{-k_{\rm s} \tau \sec \alpha}   \biggr] \bigg\}  
   \nonumber \\ 
  &  & \hspace*{0.5cm}  + \ k_{\rm h}{S_{\rm h}}~\bigg\{ 
   \biggl( {{\cos \alpha} \over {k_{\rm h}}}  + {1 \over 2}  
   \biggr) 
 \biggl[ \biggl({{\cos \alpha} \over {k_{\rm h}}} \biggr) 
   - \biggl({{\cos \alpha} \over {k_{\rm h}}}  - {1 \over 2}  
   \biggr)~e^{-k_{\rm h} \tau \sec \alpha}  \biggr] \bigg\} \ . 
\end{eqnarray}  
The derivation of the above equation will be presented elsewhere 
(Wu \& Soria, in preparation).   

In the second approximation, $B_{\rm x}(\tau)$ is assumed to have the 
same functional form as that of the first-approximation solution, i.e.\ 
\begin{equation} 
  B_{\rm x}(\tau)  \ = \  a - b_{\rm s}~e^{-k_{\rm s} \tau \sec \alpha}
                      - b_{\rm h}~e^{-k_{\rm h} \tau \sec \alpha} \ , 
\end{equation} 
where $a$, $b_{\rm s}$ and $b_{\rm h}$ are constants to be determined by 
the boundary conditions. For a semi-infinite slab opaque at optical 
wavelengths, the emergent radiation is the Laplace transform of  
$B_{\rm x}(\tau)$, i.e.\   
\begin{eqnarray}  
  I(0,\mu) & = & 
   \lim_{\tau_{\rm tot} \rightarrow \infty} \biggl[~
   \int^{\tau_{\rm tot}}_0 d\tau~ B_{\rm x}(\tau) ~ 
                     e^{- \tau /\mu}~\biggr] \nonumber  \\ 
        & = & a 
       - b_{\rm s} \biggl({{\cos \alpha} \over {k_{\rm s}}} \biggr)  
  \biggl[ {{\cos \alpha} \over {k_{\rm s}}} + \mu \biggr]^{-1} 
       - b_{\rm h} \biggl({{\cos \alpha} \over {k_{\rm h}}} \biggr) 
   \biggl[ {{\cos \alpha} \over {k_{\rm h}}} + \mu \biggr]^{-1} 
   \ . 
\end{eqnarray}   

The condition of no inward component of $I(\tau,\mu)$ at the boundary 
($\tau = 0$) implies   
\begin{equation}  
  \int_{-1}^{0} d\mu~I(0,\mu) \ = \ 0     
\end{equation}   
and 
\begin{equation}  
  \int_{-1}^{0} 
   d\mu~\mu~I(0,\mu) \ = \ 0    \ .   
\end{equation}  
With these boundary conditions, we obtain from the radiative-equilibrium 
condition   
\begin{equation}  
   a - 
   b_{\rm s} \biggl[~ 2 - \biggl({{\cos \alpha} \over {k_{\rm s}}}\biggr)~
    \ln (1 + k_{\rm s} \sec \alpha ) ~\biggr]   
 - b_{\rm h} \biggl[~ 2 - \biggl({{\cos \alpha} \over {k_{\rm h}}}\biggr)~
    \ln (1 + k_{\rm h} \sec \alpha ) ~\biggr] \ 
 = \ {1\over 2} \big( k_{\rm s} S_{\rm s} + k_{\rm h} S_{\rm h} \big)  \ .  
\end{equation}  
Similarly, the radiative-transfer equation is integrated, yielding 
\begin{eqnarray} 
     a - 2~b_{\rm s} \biggl({{\cos \alpha} \over {k_{\rm s}}}\biggr)
   \biggl[~ 1 - \biggl({{\cos \alpha} \over {k_{\rm s}}}\biggr)~
    \ln (1 + k_{\rm s} \sec \alpha ) ~\biggr]   
 - 2~b_{\rm h} \biggl({{\cos \alpha} \over {k_{\rm h}}}\biggr)
   \biggl[~ 1 - \biggl({{\cos \alpha} \over {k_{\rm h}}}\biggr)~
    \ln (1 + k_{\rm h} \sec \alpha ) ~\biggr]  &  &  \nonumber \\ 
   = \ \big( S_{\rm s} + S_{\rm h} \big)  \cos \alpha \ .  
    \hspace*{11cm} & & 
\end{eqnarray}   
Using the trivial conditions $b_{\rm s} \rightarrow 0$ 
when $S_{\rm s}\rightarrow 0$ and  $b_{\rm h} \rightarrow 0$ when 
$S_{\rm h}\rightarrow 0$, we can solve the radiative-equilibrium and 
radiative-transfer equations (12) and (13) for the 
constants $a$, $b_{\rm s}$ and $b_{\rm h}$, and obtain   
\begin{equation} 
 b_{\rm s}  \ = \  {1\over 2} k_{\rm s} S_{\rm s}  \biggl[ 
   \biggl({{\cos \alpha} \over {k_{\rm s}}}\biggr) - {1 \over 2} 
   \biggr] f_{\rm s}(\alpha)        \ , 
\end{equation}
\begin{equation} 
 b_{\rm h}  \ = \  {1\over 2} k_{\rm h} S_{\rm h}  \biggl[ 
      \biggl({{\cos \alpha} \over {k_{\rm h}}}\biggr) - {1 \over 2} 
   \biggr] f_{\rm h}(\alpha)        \ , 
\end{equation}
and 
\begin{equation} 
  a \ = \  {1\over 2} \biggl[~ k_{\rm s} S_{\rm s} 
    \biggl({{\cos \alpha} \over {k_{\rm s}}}\biggr) f_{\rm s}(\alpha) 
     + k_{\rm h} S_{\rm h} 
  \biggl({{\cos \alpha} \over {k_{\rm h}}}\biggr) 
         f_{\rm h}(\alpha)  ~ \biggr]   \ ,   
\end{equation}
where the two functions $f_{\rm s}(\alpha)$ and $f_{\rm h}(\alpha)$  
are given by 
\begin{equation} 
 f_{\rm s}(\alpha) \ = \ \biggl[~ 
   1 - \biggl({{\cos \alpha} \over {k_{\rm s}}}\biggr) 
   + \biggl({{\cos \alpha} \over {k_{\rm s}}}\biggr) 
   \biggl({{\cos \alpha} \over {k_{\rm s}}} - {1\over 2} \biggr)  
     \ln (1 + k_{\rm s} \sec \alpha ) ~\biggr]^{-1}  
\end{equation} 
and  
\begin{equation} 
 f_{\rm h}(\alpha) \ = \ \biggl[~ 
   1 - \biggl({{\cos \alpha} \over {k_{\rm h}}}\biggr) 
   + \biggl({{\cos \alpha} \over {k_{\rm h}}}\biggr) 
   \biggl({{\cos \alpha} \over {k_{\rm h}}} - {1\over 2} \biggr)  
     \ln (1 + k_{\rm h} \sec \alpha ) ~\biggr]^{-1}.  
\end{equation}  
 
We now define a hardness parameter $\xi \equiv S_{\rm h}/S_{\rm s}$ and 
a total X-ray flux $S_{\rm x} \equiv S_{\rm s} + S_{\rm h}$. 
Then, we have 
\begin{eqnarray}  
 B_{\rm x}(\tau) \ = \ {1\over 2} S_{\rm x} ~\bigg\{ 
     k_{\rm s}  f_{\rm s}(\alpha) \biggl({\xi \over {1+ \xi}} \biggr) 
      \biggl[ \biggl({{\cos \alpha} \over {k_{\rm s}}}\biggr) 
     - \biggl({{\cos \alpha} \over {k_{\rm s}}} - {1\over 2} \biggr) 
       ~e^{- k_{\rm s}\tau \sec \alpha } \biggr]  &  & \nonumber \\ 
   +   k_{\rm h} f_{\rm h}(\alpha)  \biggl({1 \over {1+ \xi}} \biggr) 
      \biggl[ \biggl({{\cos \alpha} \over {k_{\rm h}}}\biggr) 
     - \biggl({{\cos \alpha} \over {k_{\rm h}}} - {1\over 2} \biggr) 
       ~e^{- k_{\rm h}\tau \sec \alpha } \biggr] \bigg\} \ . 
    & & 
\end{eqnarray}

\begin{figure} 
\vspace*{0.25cm}  
\begin{center}
\begin{tabular}{cc}
  \epsfxsize=7.5cm \epsfbox{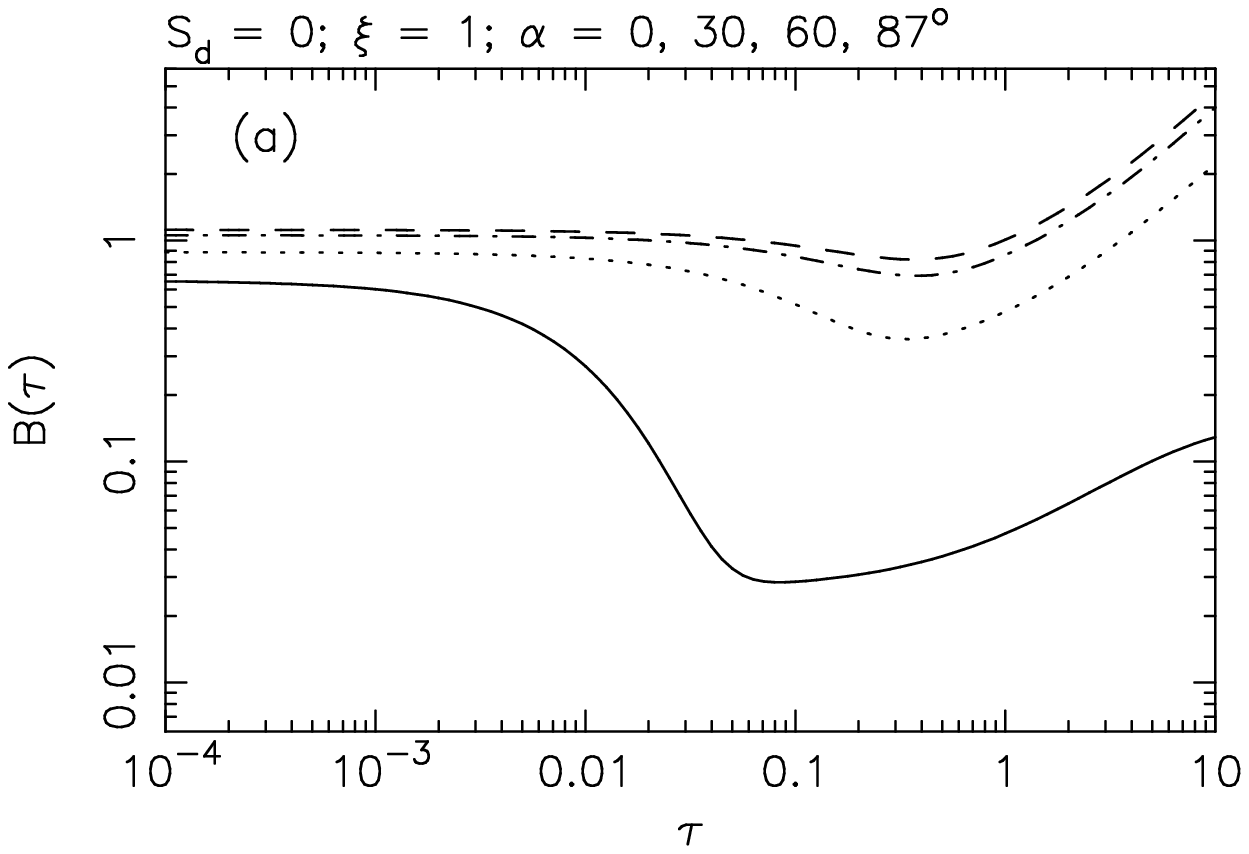} & 
  \epsfxsize=7.5cm \epsfbox{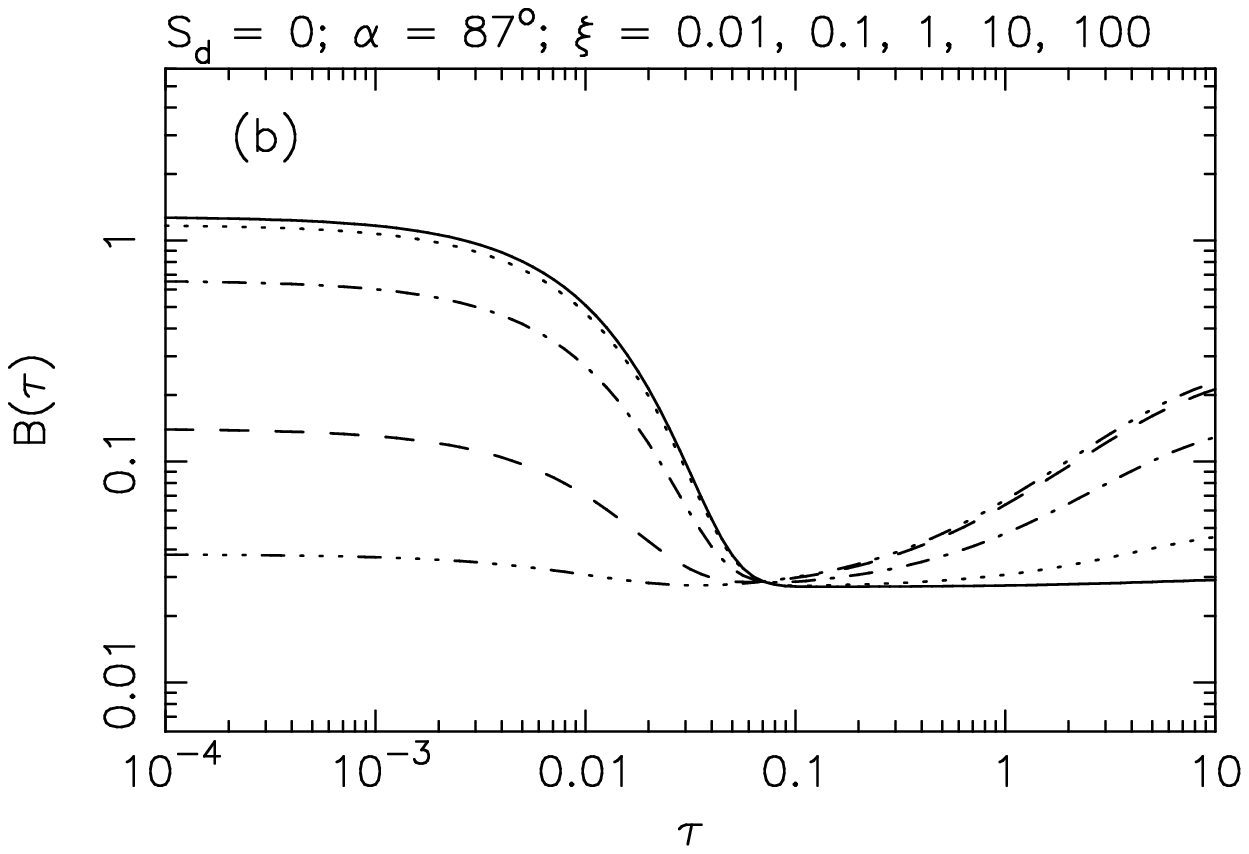} \\   
  \epsfxsize=7.5cm \epsfbox{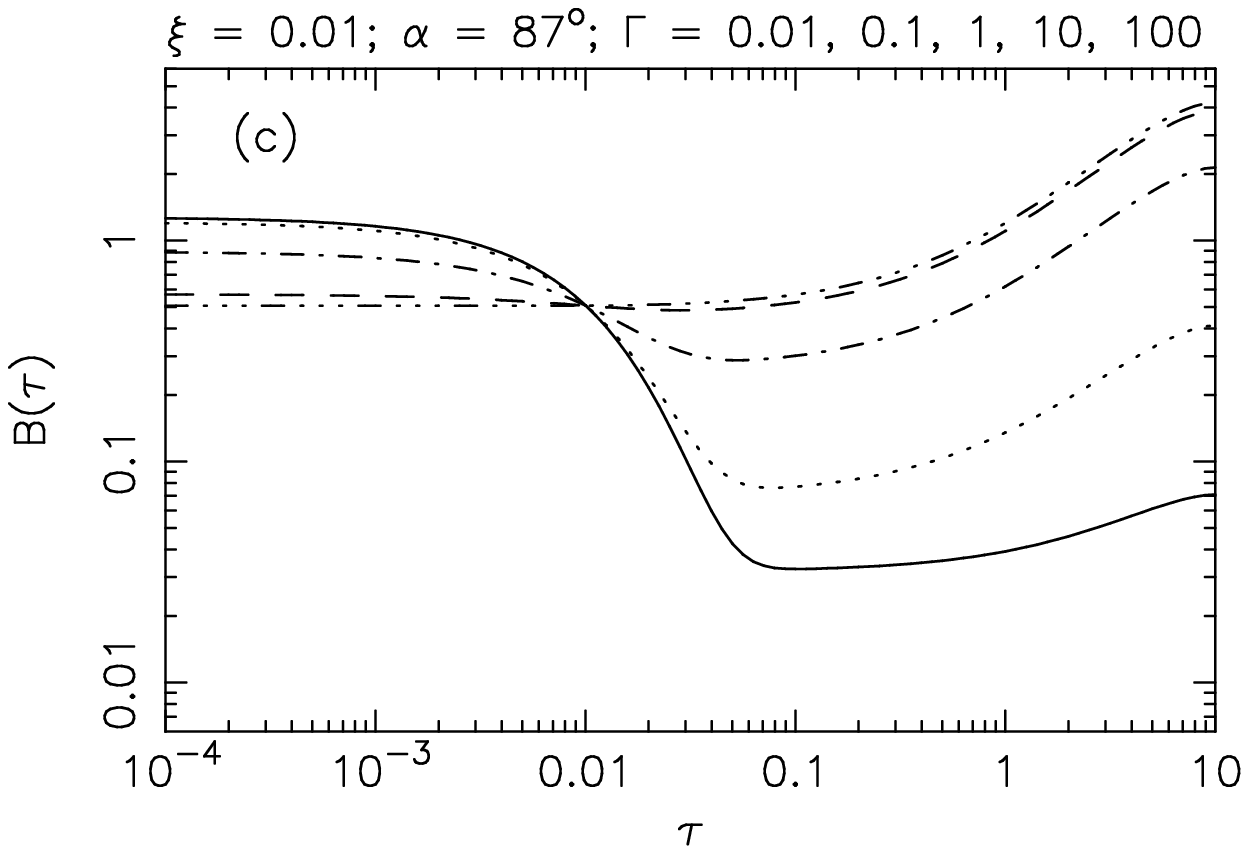} & 
  \epsfxsize=7.5cm \epsfbox{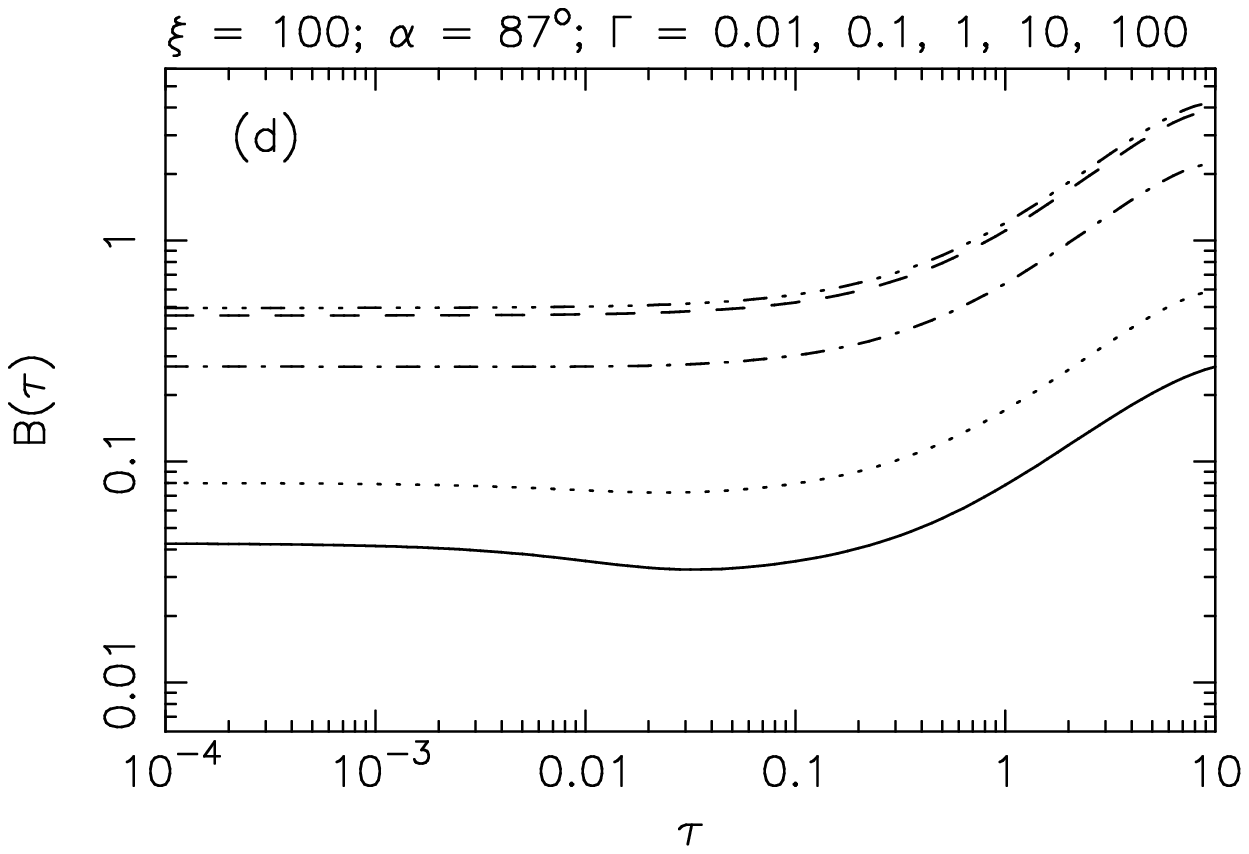} \\
  \epsfxsize=7.5cm \epsfbox{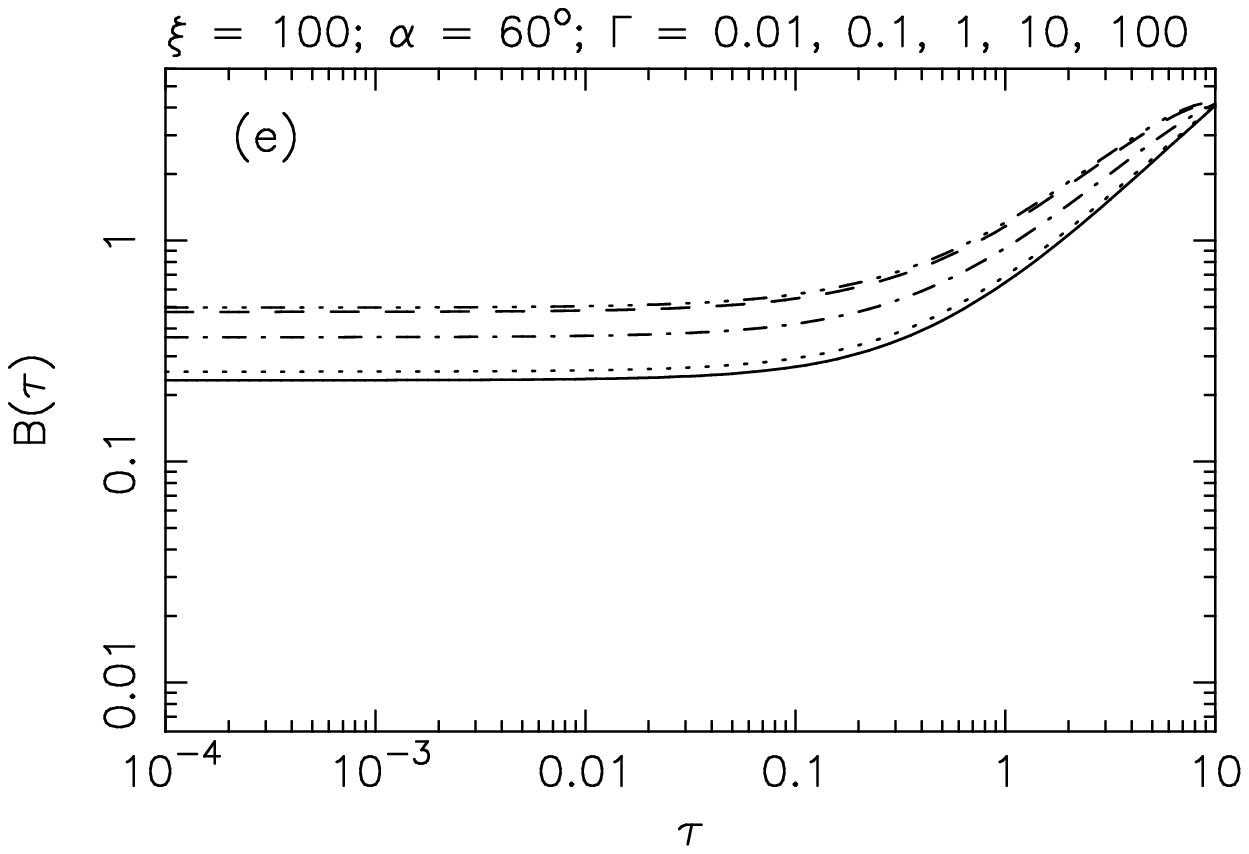} & 
  \epsfxsize=7.5cm \epsfbox{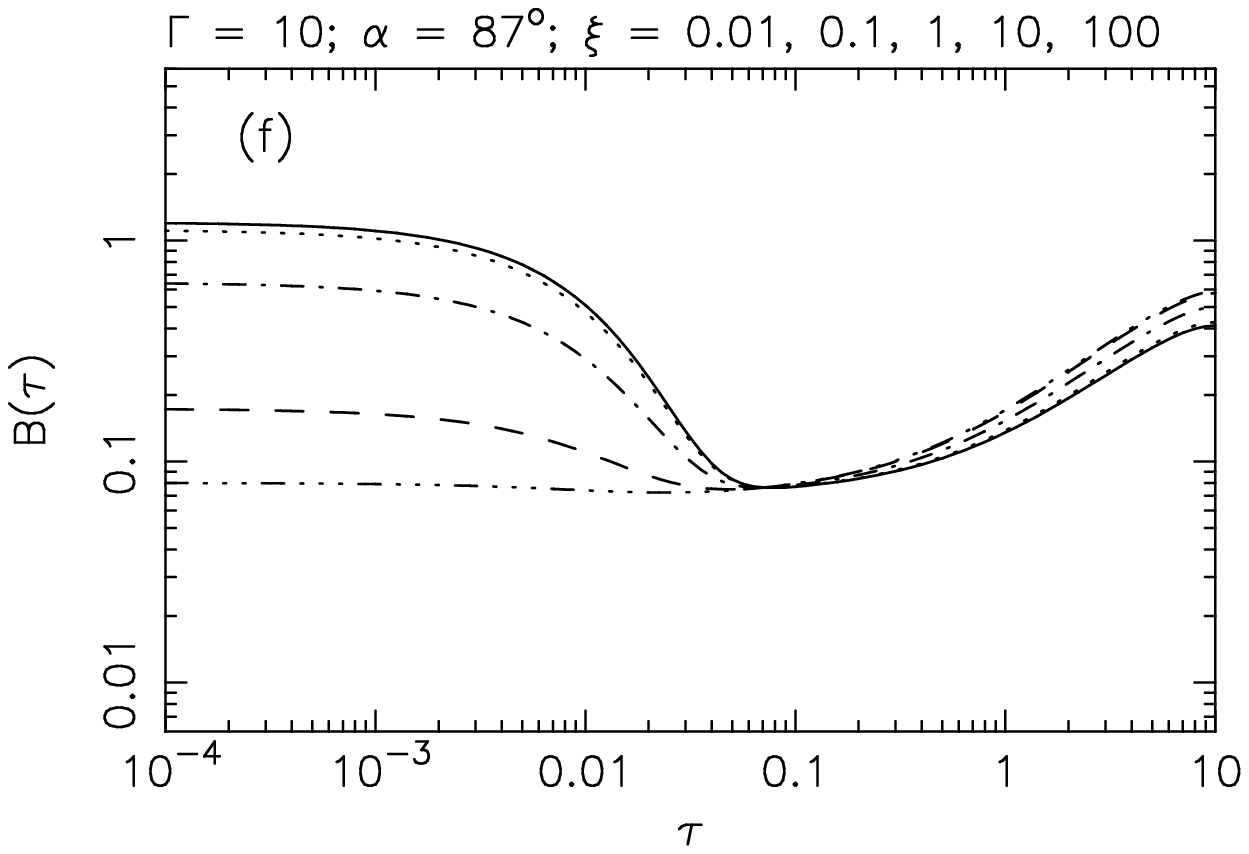} 
\end{tabular} 
\end{center}
\caption{(a) Temperature profile ($T \propto B^{1/4}$) as a function of 
  optical depth $\tau$ for X-ray incident angles $\alpha = 87^{\circ}$ 
  (solid line), $60^{\circ}$ (dotted line), $30^{\circ}$ (dash-dotted 
  line) and $0^{\circ}$ (dashed line). The viscous energy flux from 
  the disk $S_{\rm d} = 0$, 
  the X-ray incident flux $S_{\rm x} = 1$, the X-ray hardness parameter 
  $\xi =1$, and the total optical depth $\tau_{\rm tot} = 20$. 
  The two scaling parameters for the X-ray opacities $k_{\rm s} = 5.0$ 
  and $k_{\rm h} = 0.01$. 
  (b) Temperature as a function of optical depth $\tau $ for  
  X-ray hardness parameters $\xi = 0.01$ (solid line), 0.1 
  (dotted line), 1 (dash-dotted line), 10 (dashed line) and 100 
  (dash-dot-dot-dotted line). The incident angle $\alpha$ is fixed 
  at $87^{\circ}$. The other parameters are the same as in (a). 
  (c) Temperature as a function of optical depth $\tau $ for 
  non-zero viscous disk flux $S_{\rm d}$. The curves correspond to 
  illumination-strength parameters  
  $\Gamma \equiv S_{\rm x}/S_{\rm d} = 100$  (solid line), 
  10 (dotted line), 1 (dash-dotted line), 
  0.1 (dashed line) and 0.01 (dash-dot-dot-dotted line) with 
  $\xi = 0.01$ (i.e.\ strong soft X-ray illumination). We fix 
  $S_{\rm x} + S_{\rm d} = 1.0$ and $\alpha = 87^\circ$. The other 
  parameters are the same as in the previous cases. (d) Same as (c) 
  but with $\xi = 100$ (i.e.\ strong hard X-ray illumination). 
  (e) Same as (d) but with a lower incident angle $\alpha = 60^\circ$. 
  (f) Same as (b) except that here we have $\Gamma = 10$ (i.e.\  
   $S_{\rm d} > 0$). }   
\end{figure}

\subsection{The temperature-inversion layer}  

Using the plane-parallel model, we first investigate the temperature 
profile, as a function of optical depth $\tau$, in the absence of 
radiation due to viscous dissipation in the accretion disk. In this 
case the temperature profile is determined purely by the soft and hard 
X-ray irradiation. Cold disk matter has a larger opacity for soft X-rays 
than for optical radiation because of bound-free absorption. Hard 
X-rays are not easily absorbed, and electron scattering is the main 
source of hard X-ray opacity. We assume a simplified situation in 
which the an average opacity is assigned for each waveband, and  
consider the parameters $k_{\rm s} = 5.0$ and $k_{\rm h} = 0.01$ for 
the soft and hard components respectively. 

We have found that a strong 
temperature-inversion layer can be set up at the accretion-disk 
surface when the incident angle of the X-rays approaches  
grazing incidence (i.e.\ $\alpha \rightarrow 90^\circ$). 
For normal incidence, the temperature inversion is weak and 
unimportant (Fig.~7a). The temperature inversion is caused 
mainly by soft X-ray irradiation, but the more penetrating
hard X-rays tend to heat up deeper layers (Fig.~7b). 

Next we investigate the situation where viscous dissipation is also 
present, i.e.\ the temperature profile is determined jointly by viscous 
heating in the accretion disk ($S_{\rm d} > 0$) and by X-ray irradiation
($S_{\rm x} > 0$). We consider a particular situation in which the total 
flux resulting from the two processes is constant but the relative 
strength of the two processes varies. 
We show in Figure~7c the temperature profile for irradiation from a 
very soft X-ray source (with a hardness parameter $\xi = 0.01$) at 
an incident angle $\alpha = 87^\circ$. When the X-ray component 
dominates (i.e.\ $\Gamma \equiv S_{\rm x}/S_{\rm d} > 1$), a 
temperature-inversion layer is set up. When the X-rays are weak the 
temperature profile is determined by viscous heating. Figure~7d shows 
the corresponding temperature profile for the case in which the X-rays 
are hard (denoted by a hardness parameter $\xi = 100$). The temperature 
profiles are similar in spite of the different relative strength  
of the viscous heating and irradiation heating. When the incident angle 
is smaller (say $\alpha = 60^\circ$, see Fig.~7e), the effects 
due to irradiation become even less important. Although hard 
X-ray irradiation cannot significantly alter the overall temperature 
profile, it does increase the temperature uniformly at all optical depths 
--- a direct consequence of the small opacity for hard X-rays 
($k_{\rm h} \ll 1$) and of our approximation that $k_{\rm h}$ 
is depth independent. It 
is worth noting that the apparent effects at very large optical depths  
($\tau \gg 1$, see Fig.~7e) when the hard X-ray flux is much stronger 
than the viscous dissipation flux arise from the assumptions of a 
semi-infinite plane and single-sided illumination. The situation would 
be different for a finite slab with both the top and the bottom surfaces 
being illuminated; this model will be discussed elsewhere 
(Wu \& Soria, in preparation).  

We finally consider a situation in which the total X-ray flux is 
fixed at 10 times the viscous dissipation flux but its hardness 
is allowed to vary. This corresponds to strong X-ray irradiation 
with a constant bolometric X-ray luminosity and with spectral evolution. 
Our model shows that the temperature-inversion layer becomes more 
prominent as the X-ray hardness parameter $\xi$ decreases (Fig.~7f). 
Also, while the temperature at small optical depths (near the 
accretion-disk surface) has changed substantially, the temperature at 
large optical depths (in the mid-disk plane) is only modified slightly.  

In summary, the vertical temperature profile of an accretion disk is 
sensitive to soft X-ray irradiation if the X-ray flux is stronger than 
the viscous dissipation flux, but it is not significantly altered by 
hard X-ray irradiation. Soft X-ray irradiation results in the formation 
of a temperature-inversion layer on the accretion-disk surface; hard 
X-rays tend to provide uniform heating at all depths in the disk.    

\subsection{Line formation}  
 
The transfer equation for non-polarised radiation 
in a semi-infinite plane-parallel atmosphere, with complete 
redistribution and consideration of non-LTE, is  
\begin{equation} 
 {\mu} {d \over {d z}} I_\nu \ 
  = \ - (\kappa_{\rm _C}+\kappa_{\rm _L}\phi_\nu) I_\nu 
       + \kappa_{\rm _C} S_{\rm _C} 
     + \kappa_{\rm _L}\phi_\nu S_{\rm _L} 
\end{equation}  
(see Mihalas 1978). The subscripts ``$_{\rm C}$'' and ``$_{\rm L}$'' 
denote the continuum and the line-centre respectively. $\kappa_{\rm _C}$ 
and $\kappa_{\rm _L}$ are therefore the opacities of the continuum and 
line respectively. Similarly, $S_{\rm _C}$ and $S_{\rm _L}$ are 
the continuum  and line-centre source functions. $\phi_\nu$ is 
the line profile function, and $I_\nu$ is the intensity at frequency 
$\nu$. 

Without loss of generality, we simply let $\mu = 1$ ($\theta = 0^\circ$). 
As an approximation we can rewrite the transfer equation at the line-centre 
frequency as 
\begin{equation} 
 {d \over {d z}} I_{\rm _L} \ 
  = \ - (\kappa_{\rm _C}+\kappa_{\rm _L}) I_{\rm _L} 
     + j_{\rm _C} + j_{\rm _L} \ ,     
\end{equation}  
where the emissivities $j_{\rm _C}$ and $j_{\rm _L}$ are related to 
the corresponding source functions as usual. In the neighbourhood of 
the line, $\phi_\nu \rightarrow 0$, and therefore we have the transfer 
equation for the continuum    
\begin{equation} 
 {d \over {d z}} I_{\rm _C} \ 
  = \ - \kappa_{\rm _C} I_{\rm _C} 
     + j_{\rm _C} \       
\end{equation}  
(e.g.\ Tucker 1976). It follows  
\begin{equation} 
 I_{\rm _L}(z) -  I_{\rm _C}(z) \ = \   
   I_{\rm o}~[e^{-\tau(z)} - e^{-\tau_{\rm _C}(z)}] 
     + \int_0^z dz' j_{\rm _L} e^{-[\tau(z)-\tau(z')]} 
     + \int_0^z dz' j_{\rm _C}~\big[e^{-[\tau(z)-\tau(z')]}- 
     e^{-[\tau_{\rm _C}(z)-\tau_{\rm _C}(z')]}\big] \ ,      
\end{equation}  
where  $I_{\rm o}$ is the background intensity, and the total optical depth 
$\tau$ is the sum of the optical depths of the line and the continuum, 
i.e.\ $\tau  = \tau_{\rm _C} + \tau_{\rm _L}$. Generally,  
$\tau_{\rm _L} \gg \tau_{\rm _C}$. Hence, $\tau \approx \tau_{\rm _L}$. 
A line appears in emission if $I_{\rm _L} > I_{\rm _C}$, and in absorption 
if $I_{\rm _L} < I_{\rm _C}$. In a homogeneous medium,   
this is equivalent to a criterion  
$S_{\rm _L} > I_{\rm o} + {\tilde I}_{\rm _C}$ for line emission and 
$S_{\rm _L} < I_{\rm o} + {\tilde I}_{\rm _C}$ for line absorption, where 
\begin{equation} 
  {\tilde I}_{\rm _C} \ = \int_0^z dz'~ j_{\rm _C} \ . 
\end{equation} 
For systems not far away from LTE, the criterion above can be approximated 
by $B(T) > I_{\rm o} + {\tilde I}_{\rm _C}$  for line emission, and 
$B(T) < I_{\rm o} + {\tilde I}_{\rm _C}$ for line absorption. (Here,  
$T$ is the thermal temperature of the region where the line originates, 
and $B(T)$ is the local Planck function.) An inhomogeneous medium can be 
considered as a stack of homogeneous strata, and the criterion above is  
still applicable.  

As a qualitative illustration, we may consider a simple model in which 
the accretion disk consists of two strata: a central disk plane and 
a surface layer. Now, the source function of the line from the surface layer 
is denoted by $S_{\rm _L}$, and the intensity of the continuum emission by 
${\tilde I}_{\rm _C}$. The Planck function of the surface layer is $B(T)$. 
The emission from the disk beneath can be considered as the 
background $I_{\rm o}$. For emission lines to be observed, the 
line-formation region is either an extended atmosphere/corona transparent 
to the continuum or a temperature-inversion layer above the surface of 
a geometrically thin and opaque accretion disk. 

A corona can be considered as a geometrically extended surface layer. 
For certain viewing geometries (e.g. when the disk is viewed edge-on), 
the background disk emission $I_{\rm o}$ is unimportant. As the corona 
is transparent to the continuum emission, its intensity  
${\tilde I}_{\rm _C}$ is smaller than the local Planck function 
$B(T)$. The line emission is, however, opaque, implying that  
$S_{\rm _L} = B(T) > I_{\rm o} + {\tilde I}_{\rm _C}$. Hence, the 
line-emission criterion is satisfied. 

If the surface layer is geometrically thin and the disk is opaque, 
then $I_{\rm o}$ cannot be neglected. Since the temperature-inversion 
layer has higher temperatures than the disk, 
$B(T) > I_{\rm o}$. Moreover, as the temperature-inversion layer is 
optically thin to the continuum, 
$B(T) \gg {\tilde I}_{\rm _C}$. For LTE, $S_{\rm _L} = B(T)$, and the 
criterion $B(T) > I_{\rm o} + {\tilde I}_{\rm _C}$ is therefore 
satisfied, provided that the temperature of the surface layer is 
significantly higher than the temperature of the disk. For non-LTE, 
the line centre has a brightness temperature much higher than the local 
thermal temperature $T$ of the surface layer, and hence  
$S_{\rm _L} \gg B(T)~\GS~I_{\rm o} + {\tilde I}_{\rm _C}$.   

\begin{figure} 
\vspace*{0.25cm}  
\begin{center}
\begin{tabular}{cc}
  \epsfxsize=7.5cm \epsfbox{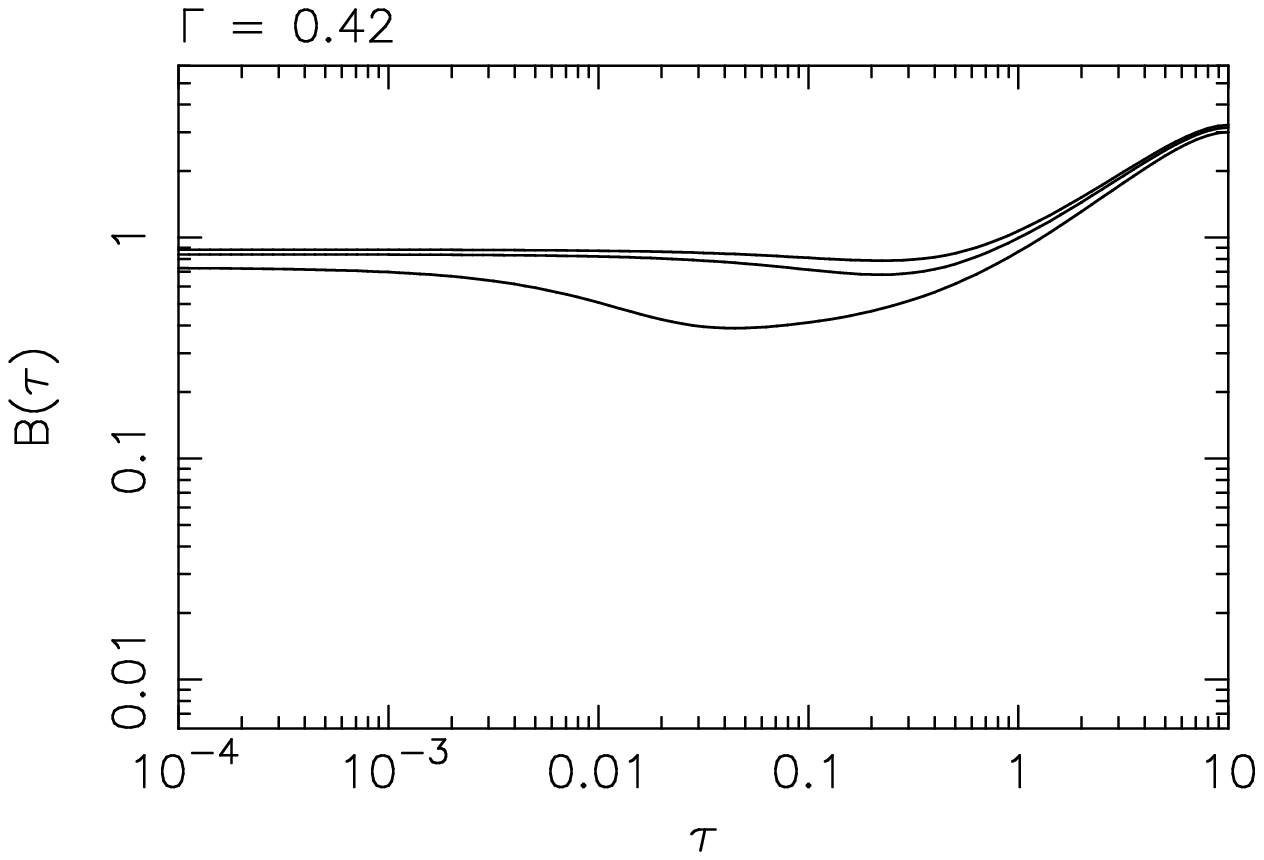} & 
  \epsfxsize=7.5cm \epsfbox{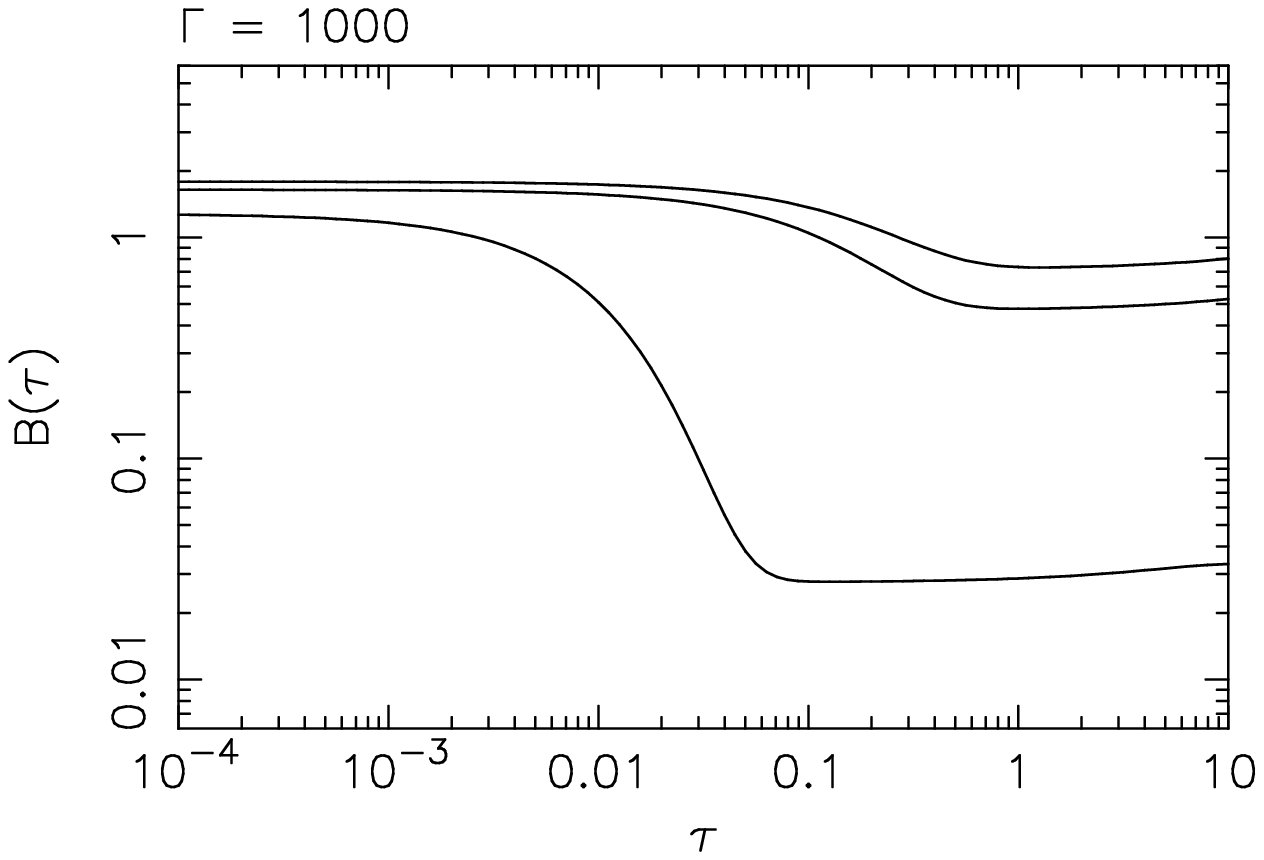} 
\end{tabular} 
\end{center}
\caption{Temperature profile as a function of optical depth $\tau$ 
 for moderate ($\Gamma \equiv S_{\rm x}/S_{\rm d} = 0.42$) and 
 strong ($\Gamma = 1000$) soft X-ray irradiation (left and right 
 panels respectively). The X-ray hardness parameter is $\xi = 0.01$, 
 the two scaling parameters for the X-ray opacities are 
 $k_{\rm s} = 5.0$ and $k_{\rm h} = 0.01$. $S_{\rm x}+S_{\rm d}$ is 
 fixed to 1.The X-ray incident angles are $\alpha = 0^\circ$, 
 45$^\circ$ and 87$^\circ$ (lines from top to bottom). }   
\end{figure}    

We now consider the temperature profiles of a region on a disk under 
moderate and strong irradiation of soft X-rays with various incident 
angles, and make use of them to investigate whether or not emission 
lines will be formed. In the case of moderate irradiation, we set 
$\Gamma$ to be 0.42 (Fig.~8, left panel). Temperature inversion 
is insignificant in this case, especially when the X-ray incident angle 
$\alpha\ \LS\ 45^\circ$, and the temperatures at the disk mid-plane 
are higher than the temperatures of the surface layers. 
For LTE the intensity of the lines from the surface layer is roughly 
the local Planck function of the layer. As    
the intensity of the emission from the disk $I_{\rm o}$ is roughly 
given by the effective temperature of the disk, there will not 
be emission lines in this case. If the line emission process is 
non-LTE, then the brightness temperature of the line is higher 
than the thermal temperatures of the surface layer and the disk, and 
hence it is possible to form emission lines.     

In the case of strong X-ray irradiation ($\Gamma =1000$) the surface 
layer is significantly hotter than the disk (Fig.~8, right panel). 
Temperature inversion is stronger for larger incident angles. As the 
Planck function of the temperature-inversion layer is greater than 
the Planck function of the disk where the optical depth is equal to 
unity, the lines are in emission.   
  
\subsection{Line-emission region of \gx339}  

In our simple plane-parallel model we have assumed parametric X-ray 
opacities that scale with the optical opacity, and  
considered a particular case, in which $k_{\rm s} = 5$ and 
$k_{\rm h} = 0.01$. We now examine whether or not these values 
are appropriate for \gx339. The effective X-ray opacity is given by 
$\kappa_{\rm x,eff} =  
   [\kappa_{\rm x}( \kappa_{\rm x} + \kappa_{\rm es})]^{1/2}$  
(Rybicki \& Lightman 1979). The X-ray opacity is 
$\kappa_{\rm x} \approx 47.9~(0.8~{\rm keV}/E)^{3}~  
 {\rm cm}^2{\rm g}^{-1}$ for $0.8~{\rm keV} < E < 8.0~{\rm keV}$  
and $\kappa_{\rm x} 
   \approx 0.17~(8.0~{\rm keV}/E)^{3}~{\rm cm}^2{\rm g}^{-1}$ 
  for $E > 8.0~{\rm keV}$ 
(e.g.\ Sincell \& Krolik 1997). The opacity due to electron scattering 
is $\kappa_{\rm es} = 0.4~{\rm cm}^2{\rm g}^{-1}$. At 
$T \sim 3 \times 10^3$--$10^4 {\rm K}$ the optical opacity is mainly 
due to atomic bound-bound transitions. An exact treatment requires the 
details of atomic processes and also the chemical composition of the 
accreting matter (see e.g.\ Collin-Souffrin \& Dumont 1990;  
Alexander, Johnson \& Rypma 1983). As an approximation, we consider   
the Rosseland mean opacity   
$\kappa \approx 0.7~(\rho/10^{-9}{\rm g~cm}^{-3})
 (T/10^4{\rm K})^{-3.5}{\rm cm}^2{\rm g}^{-1}$ 
given in Frank, King \& Raine (1992), and assume that it can be used  
to describe the opacity of the optical/UV radiation from the accretion 
disk. The soft component of the X-rays from \gx339 is a thermal black 
body with $E \sim 1$--$3 ~{\rm keV}$. If we take 
$E \approx 1.5 ~{\rm keV}$, we 
obtain a soft X-ray opacity 
$\kappa_{\rm s}\sim 7~{\rm cm}^2{\rm g}^{-1}$. The hard component is a 
power law extending between $E \sim 10$--$100 ~{\rm keV}$. If we take a 
value $E \approx 40~{\rm keV}$, we have a hard X-ray opacity 
$\kappa_{\rm h} \sim 0.02~{\rm cm}^2{\rm g}^{-1}$. The optical opacity 
is $\sim 1~{\rm cm}^2{\rm g}^{-1}$. Our choices of 
$k_{\rm s}$ and $k_{\rm h}$ are therefore appropriate.   

\begin{figure} 
\vspace*{0.25cm}  
\begin{center}
\begin{tabular}{ccc}
  \epsfxsize=5.5cm \epsfbox{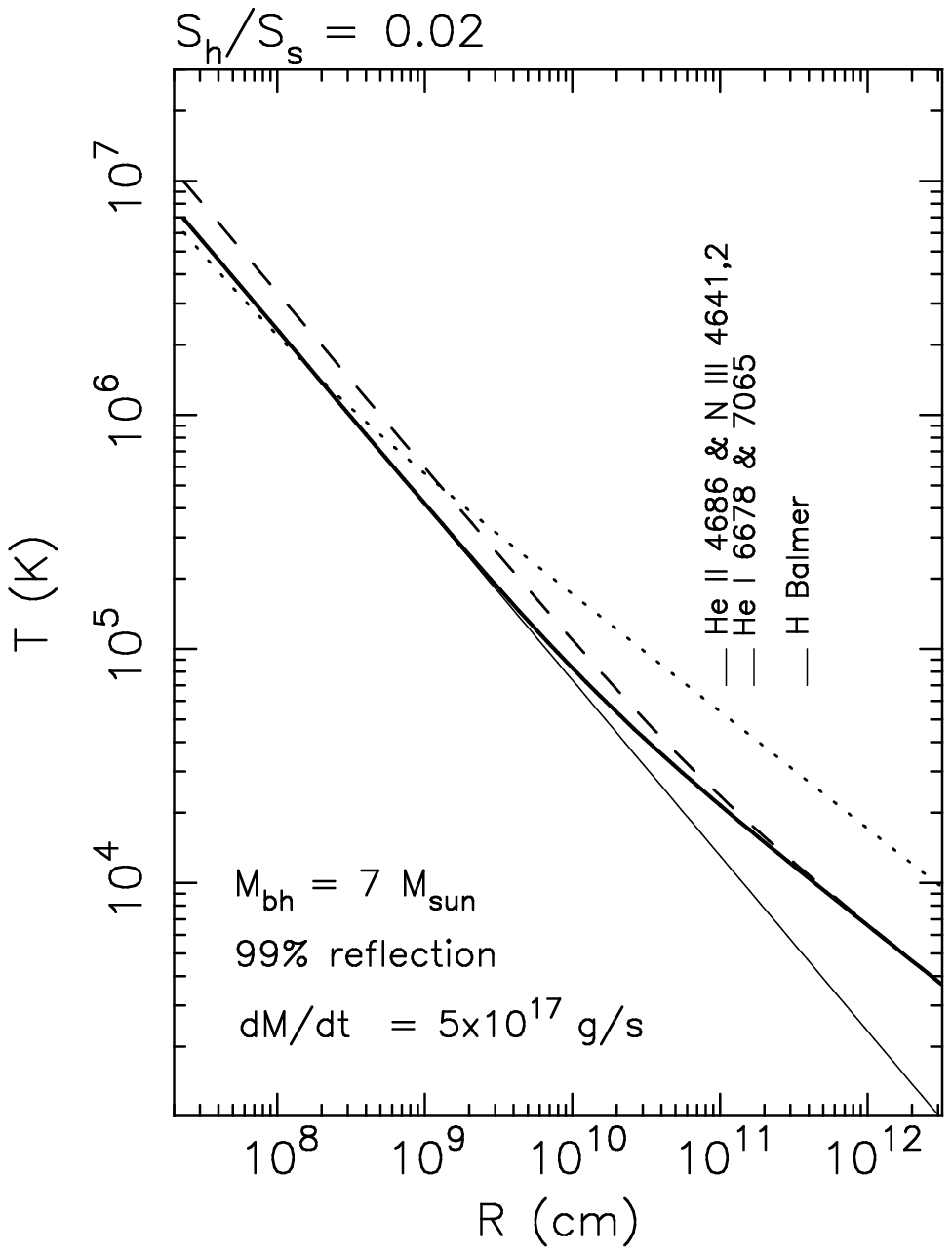} & 
  \epsfxsize=5.5cm \epsfbox{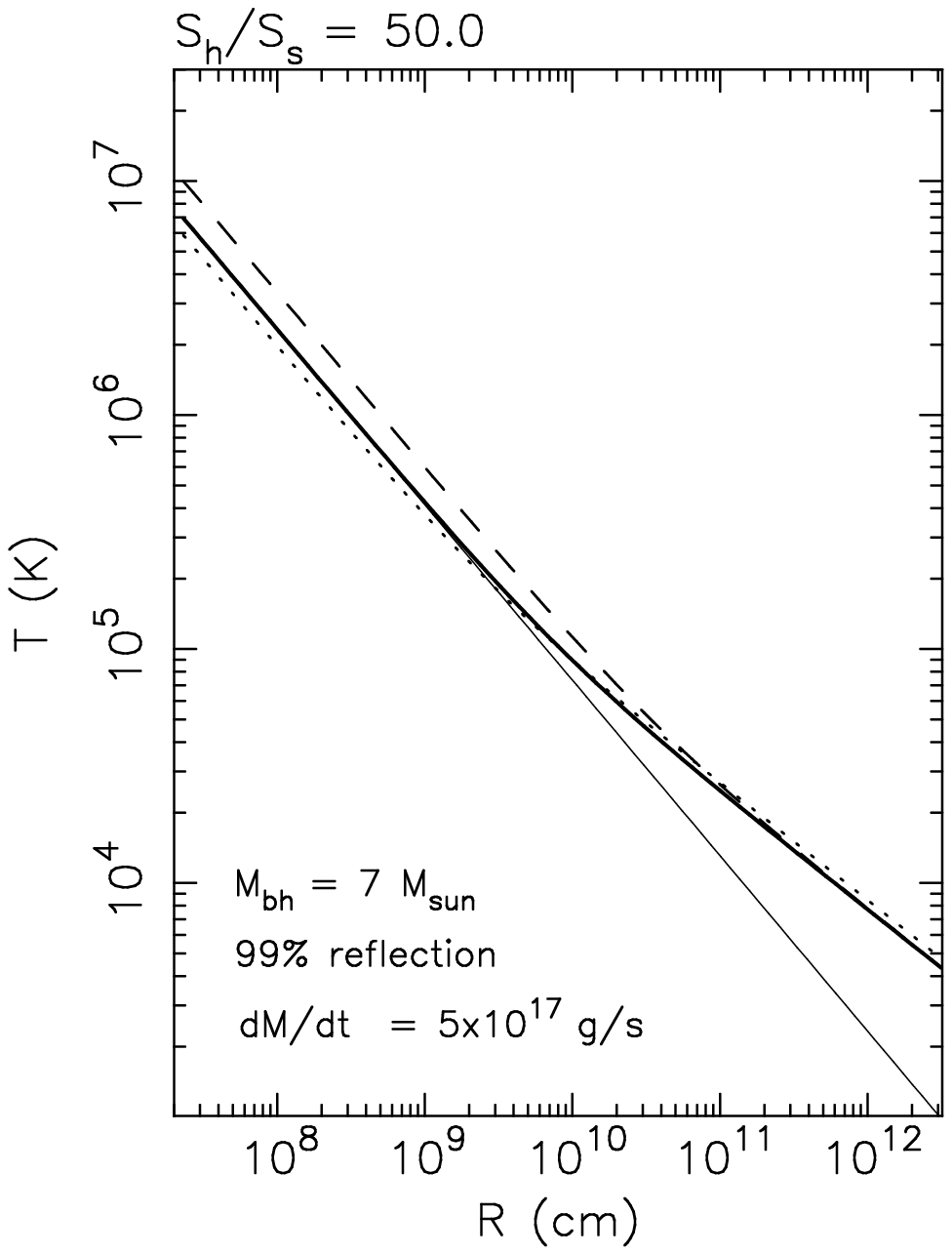} & 
  \epsfxsize=5.5cm \epsfbox{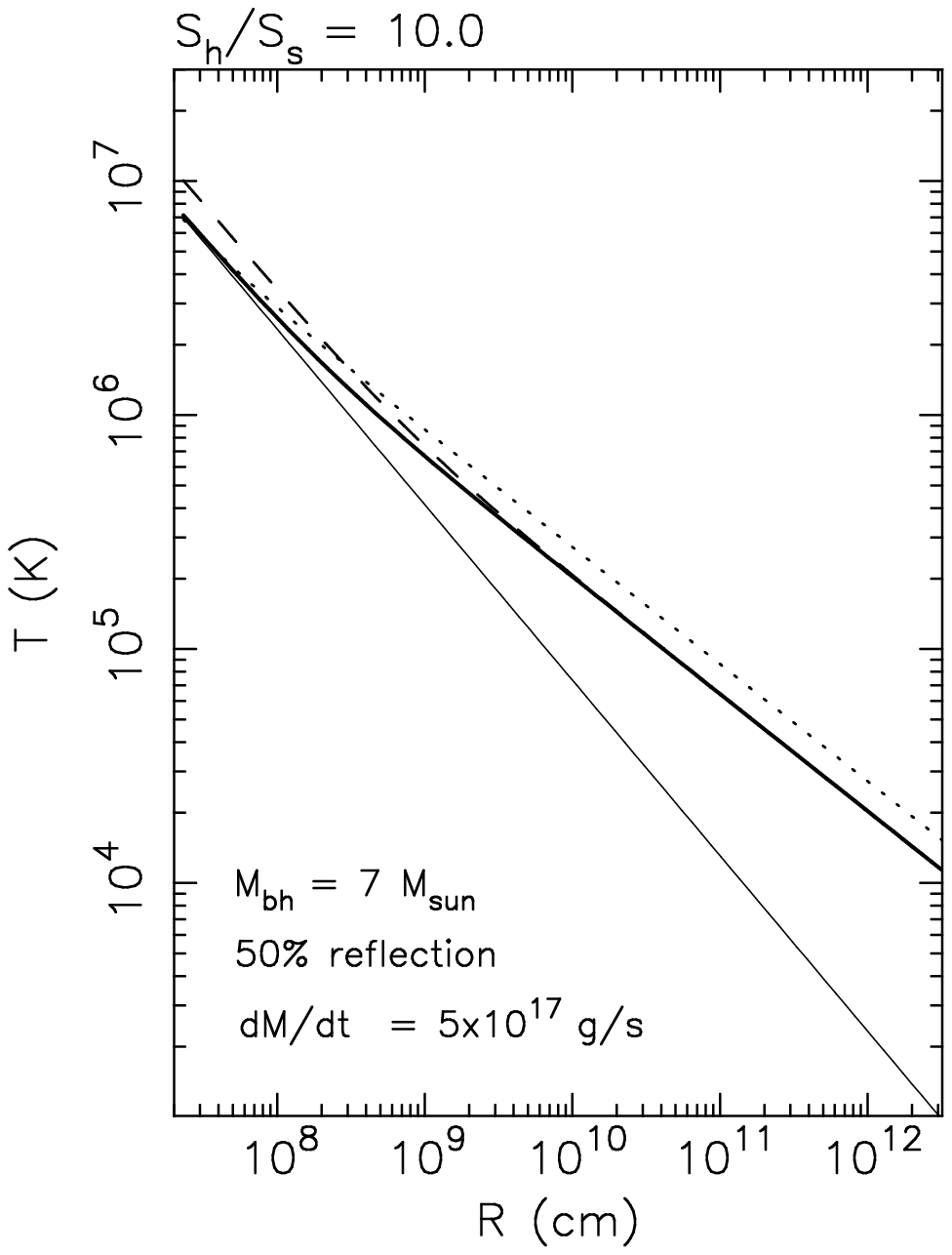} 
\end{tabular} 
\end{center}
\caption{The temperature structures of an irradiatively heated 
  accretion disk calculated from our plane-parallel model. 
  The surface temperature (at $\tau = 0$) of the disk is represented 
  by the dotted line; the effective temperature (at $\tau = 2/3$) by 
  the thick solid line; and the temperature at $\tau = 10$ by the 
  dashed line. We have assumed a Shakura-Sunyaev disk with a viscosity  
  parameter $\alpha_{\rm vis} =1$. For comparison we also show the 
  effective temperature (thin solid line) of the same Shakura-Sunyaev 
  disk without X-ray heating. The mass of the black hole 
  is 7~M$_\odot$; the mass-accretion rate is 
  $5 \times 10^{17}~{\rm g~s}^{-1}$; and the incident angle of the  
  X-rays is fixed to be $87^\circ$. In the left panel, the disk is 
  irradiated by soft X-rays with a hardness ratio $\xi = 0.02$, and  
  99\% of the X-rays are reflected. We also mark the radii 
  at which the Balmer, He\,{\scriptsize I} $\lambda$\,6678, 
  He\,{\scriptsize I} $\lambda$\,7065, 
  He\,{\scriptsize II} $\lambda$\,4686 and 
  N\,{\scriptsize III} $\lambda \lambda$\,4641,4642 lines are emitted, 
  inferred from the peak separation observed from \gx339 in 1998. 
  In determining the location of the line-emission region, we assume 
  an orbital inclination $i$ of 15$^\circ$, a constraint imposed by 
  the 14.8-hr period (\S3 and Callanan et\,al.\ 1992), and that 
  the lines are emitted from a geometrically thin, Keplerian accretion 
  disk (see Fig.~2). If the period is longer or the orbital 
  inclination is larger, the line-emission regions will be at some  
  larger radii (See Fig.~2). Balmer lines are emitted at 
  temperatures $\sim (1 - 2) \times 10^4$~K; 
  He\,{\scriptsize I} lines at $\sim (2 - 4)  \times 10^4$~K; 
  and He\,{\scriptsize II} lines at 
  $\sim (5 - 10) \times 10^4$~K (see e.g.\ 
  Cox \& Tucker 1969; Dopita \& Sutherland 1996; Osterbrock 1989). The 
  Bowen N\,{\scriptsize III} $\lambda \lambda$\,4641,4642 lines are 
  formed in regions where He\,{\scriptsize II} $\lambda$\,304 photons 
  are abundant. As shown, 
  the temperatures at the inversion layer are consistent with 
  the temperatures at which these lines are emitted. In the middle panel, 
  the accretion disk is irradiated by X-rays with a 
  hardness ratio of $\xi = 50$, and 99\% of the X-rays 
  are reflected. In the right panel, the 
  hardness ratio $\xi = 10$, and only 50\% of the X-rays are reflected. 
}   
\end{figure}    

In a previous section, we have proposed that the double-peaked lines 
seen in \gx339 during the high-soft state  originate from a  
temperature-inversion layer above an irradiatively heated, 
Keplerian-like accretion disk. Now, we are going to use the 
plane-parallel model to test whether this scenario is consistent with 
our 1998 observations.   

For black-hole binaries, most of the X-rays are emitted from a small 
region within 1000 Schwarzschild radii from the central black hole. The 
linear size of the X-ray emission region is therefore  
$< 10^9 {\rm cm}$. For a system with a period greater than 12~hr, 
the radius of the accretion disk would extend beyond $\sim 10^{11} 
{\rm cm}$, and hence the incident angle of the X-rays on the accretion 
disk is close to $90^\circ$.  

In our calculation, we consider a standard Shakura-Sunyaev disk around 
a 7-M$_\odot$ black hole under irradiation of X-rays at an incident angle 
of 87$^\circ$. We assume a viscosity parameter of $\alpha_{\rm vis} =1$  
for the accretion disk, and the X-ray luminosity from the central 
(point) source given by $L_{\rm x} = \eta {\dot M}c^2$, where $\dot M$ 
is the mass-accretion rate and $c$ the speed of light. We adopt an 
efficiency parameter $\eta = 0.1$ following Frank, King \& Raine (1992), 
and a mass-accretion rate of $5 \times 10^{17}~{\rm g~s}^{-1}$. Taking 
into account that most of the soft X-rays are reflected at grazing 
incidence, a reflective efficiency of 99\% 
(see de Jong, van Paradijs \& Augusteijn 1996) is assumed. 

In the left panel of Figure 9 we show the temperature structure 
of the accretion disk illuminated by soft X-rays with a hardness ratio 
$\xi = 0.02$. Strong temperature inversion occurs at radii 
$R\ \GS\ 10^9$~cm, with surface temperatures (at $\tau = 0$) 
about $10^4 - 10^5$~K at $R \sim 10^{11}$~cm. The effective temperature  
(evaluated at $\tau \approx 2/3$) is about a factor of $2 - 3$ 
lower. We also show the locations of the line-emission regions deduced 
from the peak separations of the lines during the 1998 
high-soft state, for an assumed orbital inclination of 
$i = 15^\circ$ (cf. Fig.~2). The temperature distribution resulted 
from our model calculation is in agreement with the temperatures of the 
formation regions of double-peaked hydrogen and helium lines.   

The middle panel of Figure 9 shows the same accretion disk but under 
hard X-ray irradiation. The hardness ratio is $\xi = 50$. The soft 
component of the X-rays is weak, and is insufficient to 
create a prominent temperature inversion. For LTE processes, the 
lines emitted from the disk should be suppressed. It is worth noting 
that although the composition of the X-rays is very different in 
the two cases, the effective temperature of the disk is similar. 
While the emission lines are formed in the temperature-inversion layer, 
the continuum emission is characterised by the effective temperature. 
One therefore expects that the optical brightness in the two cases will   
not differ significantly. Thus, our model shows that the optical continuum 
is relatively insensitive to the X-ray hardness, as observed in \gx339, 
provided that other factors do not alter significantly when the X-ray 
hardness varies.  

For comparison we also consider an accretion disk irradiated by moderately  
hard ($\xi =10$) X-rays and assume that only 50\% of the X-rays are 
reflected (Fig.~9, right panel). As shown, the entire accretion disk 
is severely heated, because the energy flux deposited by the X-rays is 
larger than the energy flux generated by viscous heating in the disk.   
There is a temperature inversion at the disk surface because of a 
non-negligible soft component, but it is not as strong as the case where 
the X-rays have an overwhelmingly strong soft component. Under such severe 
irradiative heating by the hard X-rays, the structure of the accretion disk 
should deviate significantly from the structures deduced from the 
Shakura-Sunyaev prescription. The radiative-transfer and hydrodynamics 
equations must be solved simultaneously in order to obtain a self-consistent 
temperature structure of the accretion disk. (This is beyond the scope 
of the present paper.) Nevertheless, we have shown here that the accretion 
disk responds very differently to X-ray irradiation of different hardness. 
  
We suspect that a prolonged low-hard state causes parts of the accretion 
disk in \gx339 to be heated up. The outer disk inflates to become 
an extended isothermal atmosphere/corona, and the temperature-inversion 
layer is quenched in this region. The X-ray heating may also drive a 
disk wind. The Balmer lines, which are emitted from the extended 
atmosphere/corona or wind, lose their Keplerian-disk signature and become 
single-peaked (cf.\ Murray et\,al.\ 1995). 
The He\,{\scriptsize II} $\lambda$\,4686 line is emitted 
at a smaller radial distance from the compact object, where the accreting  
material is still relatively confined to the disk plane, and the residual 
soft component of the X-rays causes a local temperature inversion.  
He\,{\scriptsize II} $\lambda$\,4686 may therefore retain its double-peaked 
profile. 

Although the plane-parallel model considered here produces results 
consistent with our spectroscopic observations of 
\gx339 in the 1998 high-soft state, we are aware of its limitations. 
In particular, non-LTE effects are not considered. Moreover, we 
have assumed a ``semi-grey'' approach, and the atomic processes that 
determine the radiative 
equilibrium and govern the radiative transfer are not treated  
explicitly. The plane-parallel model is sufficient for the purpose 
of a qualitative illustration, but a proper study should treat all 
these effects in more detail.  

Double-peaked lines are also observed 
from some black-hole candidates in quiescence, for example 
A0620-00. However, we suspect that the mechanism of emission is very 
different in the two states. 
We showed in \S5.3 that line emission can occur in two 
physical conditions: either from a temperature-inversion layer 
on the surface of optically thick gas, or from optically thin gas;
in the former case, the lines are superimposed on the continuum emission.
In the case of GX339-4 in outburst, 
the double-peaked optical emission lines come 
from the irradiated surface of an optically thick disk. 
In the case of A0620$-$00 and similar systems in quiescence 
(or more generally at very low accretion rate and X-ray luminosity), 
the accretion disk is probably optically thin, hence it is hot.
In this case, emission lines do not require a temperature-inversion 
layer to be formed, hence they are not caused by soft X-ray irradiation. 
It is worth noting that \gx339 in quiescence (Orosz 2000, private 
communication) also shows Balmer emission while the $V$ brightness 
is at least five magnitudes lower than in the soft and hard X-ray states,
suggesting that the accretion disk is optically thin in the continuum.

\section{Conclusions}  

Our observations have shown that the profiles of the emission lines 
from \gx339 are different in the high-soft and low-hard X-ray 
spectral states. In particular, the hydrogen Balmer lines have 
single-peaked profiles in the low-hard state but double-peaked 
profiles in the high-soft state. The 
He\,{\scriptsize II} $\lambda$\,4686 line is doubled-peaked in both 
states. Our interpretation is that 
soft and hard X-rays have different heating effects on the accretion 
disk and, as a result, the line-formation mechanisms in the two X-ray 
states are different. We have constructed a simple 
plane-parallel model to illustrate the differences in the two cases. 
Our model has shown that a strong temperature-inversion layer 
can be formed when the X-rays are soft and they illuminate 
the disk at an angle approaching grazing incidence. If the double-peaked 
lines that we observed during the high-soft state are emitted from 
the temperature-inversion layer on a geometrically thin, opaque, 
Keplerian accretion disk, we deduce that \gx339 is a low-inclination 
system. The orbital inclination will be about 15$^\circ$, for an orbital 
period of 14.8~hours.

\section{Acknowledgements} 

We thank the referee for directing our attention to the importance of 
the effects of X-ray reflection. KW acknowledges the support from 
the Australian Research Council (ARC) through an Australian Research 
Fellowship and an ARC grant. RS acknowledges the support of the 
Research School of Astronomy and Astrophysics, 
Australian National University, and of 
the Research Centre for Theoretical Astrophysics, 
University of Sydney.
RWH acknowledges the support from the ARC through a research grant.

\end{document}